\documentclass[noeprint,twocolumn,showpacs,amsmath,amssymb,sort&compress,floatfix,superscriptaddress,pre,aps]{revtex4-1}
\usepackage{graphicx}
\usepackage{dcolumn}
\usepackage{bm}
\usepackage{graphicx, amsmath, xcolor}
\newcommand{\lbar}{\Bar{\ell}}

\newcommand{\ea}{\sigma_{a}}

\newcommand\rv[1]{{\color{black} #1}}

\newcommand{\ucsb}{Department of Physics, University of California Santa Barbara, Santa Barbara, CA 93106, USA}
\newcommand{\bmse}{Interdisciplinary Program in Quantitative Biosciences, University of California Santa Barbara, Santa Barbara, CA 93106, USA}
\newcommand{\bioe}{Departments of Bioengineering and of Mechanical Engineering, University of California Santa Barbara, Santa Barbara, CA 93106, USA}

\begin{document}

\title{Spontaneous and Induced Oscillations in Confined Epithelia}

\author{Toshi Parmar}
\affiliation{\ucsb}
\author{Liam P. Dow}
\affiliation{\bmse}
\author{Beth L. Pruitt}
\affiliation{\bioe}
\affiliation{\bmse}
\author{M. Cristina Marchetti}
\affiliation{\ucsb}
\affiliation{\bmse}

\date{\today}

\begin{abstract}
    The feedback between mechanical and chemical signals plays a key role in controlling many biological processes and collective cell behavior. Here we focus on the emergence of spatiotemporal density waves 
    in a one-dimensional "cell train." 
    Combining a minimal theoretical model with 
     \emph{in vitro} experiments on MDCK epithelial cells confined to a linear pattern, we examine the spontaneous oscillations driven by  feedback between myosin activation and  mechanical deformations, as well as their effect on the response of the tissue to externally applied deformations.
 We show that the nature and frequency of spontaneous oscillations is controlled by the size of the cell train, with a transition from size-dependent standing waves to intrinsic spontaneous waves at the natural frequency of the tissue. The response to external boundary perturbations exhibit a resonance at this natural frequency, providing a possible venue for inferring  the mechanochemical couplings that control the tissue behavior from rheological experiments.
\end{abstract}
\maketitle

\section{Introduction}
Epithelial tissues display collective cell dynamics and spontaneous spatiotemporal oscillations in many settings~\cite{Petrolli2021OscillationsMigration}. Examples include coordinated cell motion in zebrafish fin-regeneration~\cite{DeLeon2023MechanicalTailfin}, contraction pulses in \emph{T. Adherens}~\cite{Armon2021}, immune cell migration~\cite{Gong2023Chemo-mechanicalPodosomes}, pulse ratchets in epithelial morphogenesis~\cite{Staddon2019MechanosensitiveMorphogenesis}, and epithelial migration on patterned substrates~\cite{Fang2022ActiveSurfaces}. Self-sustained oscillations are seen in tissues undergoing unconstrained expansion~\cite{Serra-Picamal2012MechanicalExpansion, Tlili2018CollectiveVelocity, Boocock2020TheoryMonolayers} and in
    confined epithelia~\cite{Deforet2014ARTICLEConfinement, Peyret2019SustainedSheets, Petrolli2019Confinement-InducedModes}, where the period and wavelength of these spontaneous spatiotemporal waves are often set by the confinement dimensions. Confined geometries are not realized only  \emph{in vitro}, but also occur frequently \emph{in vivo}.  An example is the migration of border cell clusters in \emph{Drosophila} oocytes~\cite{Cai2014MechanicalMigration}. 
    
  Theoretical models have described  epithelia  as active elastic materials~\cite{Banerjee2015PropagatingExpansion, Notbohm2016CellularMotion, Banerjee2019ContinuumMigration, Alert2020PhysicalMigration, Vincent2015ActiveMonolayer, Safa2024ActiveMechanics}, with
  mechanochemical feedback between tissue deformation and active stresses induced by  activation of proteins like myosins.
    These models have been able to reproduce observed spatiotemporal patterns \cite{Bailles2022MechanochemicalTissues}, but it is generally difficult to extract model parameters from experiments.
    Recent experiments have  started to probe
    the response of epithelia to external mechanical perturbations~\cite{Khalilgharibi2019StressCortex, Sadeghipour2018, Bodenschatz2022EpithelialStress, Pullen2021SkinEpithelia, Harris2012CharacterizingMonolayers, Borghi2012}. Experiments have shown that stretched cell monolayers can stress stiffen~\cite{Kinoshita2020MechanicalEmbryogenesis}, fluidize~\cite{Trepat2007UniversalCell} or fracture~\cite{Harris2012CharacterizingMonolayers}, demonstrating the need for a better understanding of the interplay between internally driven active stresses and externally imposed deformations in controlling the rheology of biological tissue.

In this study, we consider a minimal theoretical model of epithelium as an active elastic material that incorporates the mechanochemical feedback between mechanical deformations and phosphorylated myosins. Using this model, we examine the emergent dynamics of  a tissue confined to a one-dimensional (1-D) geometry, as well as its response to externally imposed oscillatory deformations. We validate our predictions with observations in an \textit{in vitro} experimental system of  MDCK epithelial cells confined to a 1-D collagen type I pattern. \rv{The study of cell collectives in confined and patterned geometries has become a very useful tool for probing the adaptive mechanics of living tissue. In particular,  many cell properties have been explored by confining cells to both unconstrained~\cite{Vercruysse2024NatureClusters,Rossetti2024OptogeneticMigration} and finite-length~\cite{Petrolli2019Confinement-InducedModes} 1-D geometries.
}

We show that the confined tissue exhibits self-sustained spontaneous oscillations arising from the feedback between myosin-driven contractility and strain-driven myosin activation, as found in previous work~\cite{Boocock2020TheoryMonolayers}. These oscillatory modes are contraction/expansion waves. If the length of the tissue is smaller than a critical length, the frequency and wavelength of oscillations are controlled by the tissue size. When the tissue length exceeds this critical value, the frequency and wavelength of spontaneous sustained oscillations are regulated by intrinsic properties of the tissue and can be obtained by a simple linear stability analysis. This crossover between size-dependent standing waves and intrinsic oscillations has been observed before and has been understood via a self-propelled Voronoi model of the tissue as arising from the cells' tendency to align their polarization with their velocity~\cite{Petrolli2019Confinement-InducedModes}. Here we show that a similar behavior also arises in our theoretical model of an isotropic tissue from the feedback between myosin activation and mechanical stresses, without an imposed  alignment mechanism. In the experimental realization of our cell train confinement may induce local cell polarization,
\rv{which has  been shown to play an important role in cell migration in confined and complex environments~\cite{Vercruysse2024NatureClusters}.} Our model demonstrates, however, that explicit alignment is not required to set up oscillations. \rv{On the other hand, contact regulation of cell polarity and locomotion reorganization are undoubtedly important to capture the interaction of cell groups with physical boundaries~\cite{Vercruysse2024NatureClusters}.}  

We then study the response of the tissue to localised step-strains and compare the response of the model to \emph{in vitro} experiments on MDCK cells confined to a linear geometry by substrate patterning. \textit{In vitro} the 1-D cell collective shows a preference for moving in the direction of the mechanical stimulus provided by the applied step-strain  on a similar timescale as in our theoretical model. By comparing the model to experiments we find that the response to a short-lived stimulus  depends on the phase of spontaneous oscillations of the epithelium and that a  mechanical perturbation applied don longer time scales is needed to probe force transmission. We then examine the response of the tissue to an oscillatory driving force applied at the boundaries. We show that in this case, the tissue oscillates at the externally applied frequency as well as at its intrinsic oscillation frequency, and shows  resonance  when the two frequencies match. We show that one can extract the mechanochemical coupling strength in this system from measurement of the penetration length.

The organisation of the rest of the paper is as follows: we first provide model details in Section II followed by the study of emergent spontaneous oscillations in confined and unconfined tissues and the oscillation properties in Section III. In Section IV we \rv{introduce the experimental setup and} compare the results of step-strain perturbation in model and experiments. In Section V we explore the response to oscillatory boundary forces and conclude our results in Section VI.

\section{Model Details}

We construct a minimal model of an epithelial monolayer that accounts for the mechanochemical feedback between myosin activation and cell elasticity. The model is motivated by experiments on 1-D cell trains confined by patterning adhesive proteins \cite{Chen1998MicropatternedFunction, Tang2012AGels, Moeller2018ControllingPatterning} and inspired by related models previously considered in the literature~\cite{Banerjee2015PropagatingExpansion, Boocock2020TheoryMonolayers, Armon2021}. 

\begin{figure}[!ht]
    \centering  
    \includegraphics[width=0.45\textwidth]{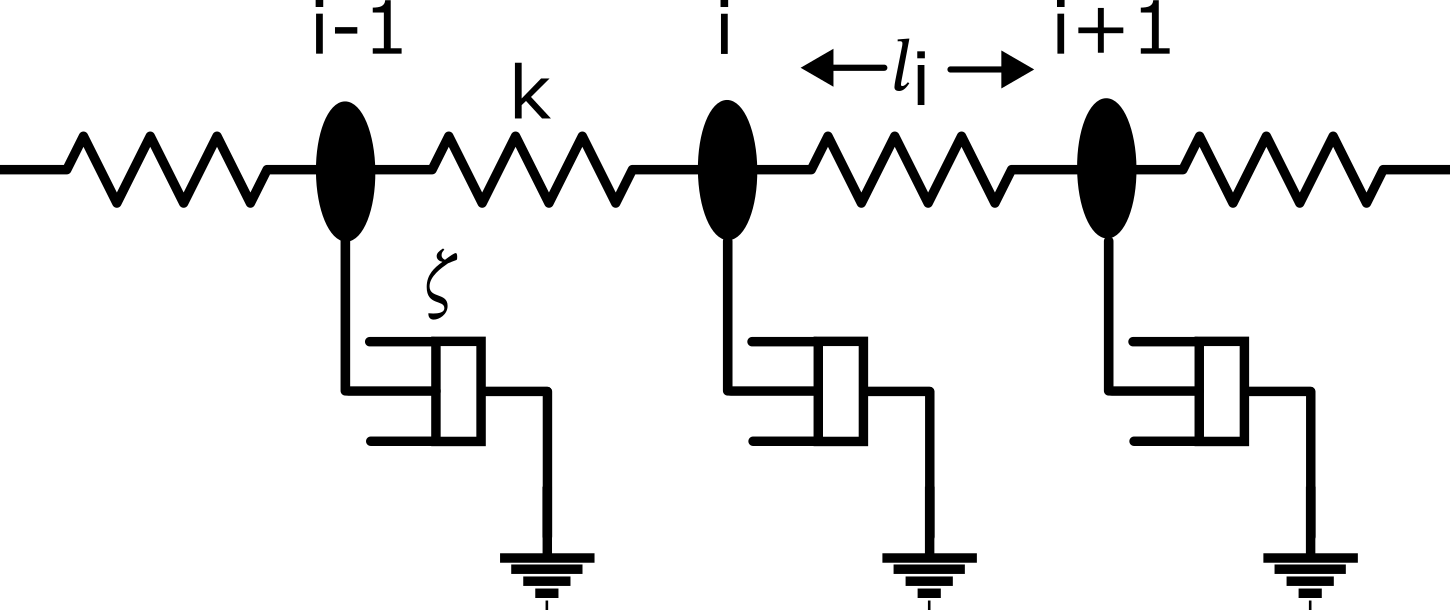}
    \caption{Schematic of our model of a $1D$
    chain of active elastic cells with stiffness $k$ and friction $\zeta$. The black ovals denote the cell vertices which experience traction from the substrate.}
    \label{fig:1D_chain}
\end{figure}

We consider a linear chain of $N$ cells bounded by vertices labelled by $i=0,1, \cdots, N$ (Fig.~\ref{fig:1D_chain}). The unstretched tissue is confluent and has total length $L$, with $\lbar=L/N$ a mean cell length that we take as unit of length.
The $i$-th cell is bounded by vertices at $r_i$ and $r_{i+1}$ with internal tension $T_i$. 
The overdamped dynamics of each vertex is controlled by force balance, given by
\begin{equation}
    \label{eq:balance}
    \zeta\dot r_i=T_i-T_{i-1}\;,
\end{equation}
where $\zeta$ is a friction and $T_i$ is the tension in cell $i$, 
given by
\begin{equation}
\label{eq:tension}
    T_i=k\left(\ell_i-\ell_{i0}\right)\equiv k\lbar\epsilon_i\;,
\end{equation}
with $k$ the stiffness of the cell, $\ell_i(t)=r_{i+1}(t)-r_i(t)$ the length of cell $i$ at time $t$,  $\epsilon_i(t)=(\ell_i(t)-\ell_{i0}(t))/\lbar$  the strain and $\ell_{i0}(t)$ the rest length of cell $i$ at time $t$. We define the displacement of each vertex as, $u_i(t)=r_i(t)-i\lbar$.

Cells in tissue are under tension to balance intracellular contractile forces and  traction forces exchanged with the substrate. 
The rest length $\ell_{i0}(t)$ of each cell  is defined as the length where the tension $T_i$ vanishes. It is itself a dynamical variable because it can be changed by myosin recruitment. \rv{Previous studies on both confined \cite{Deforet2014ARTICLEConfinement, Notbohm2016CellularMotion} and expanding \cite{Serra-Picamal2012MechanicalExpansion} epithelia have shown the essential role of myosin in driving sustained spontaneous oscillations by demonstrating that the addition of myosin inhibitors leads to the loss of oscillatory behavior.} To incorporate this mechanochemical feedback, we add equations describing the dynamics of the rest length and the  concentration $c_i(t)$  of phosphorylated myosin in the $i$-th cell at time $t$ as
\begin{align}
    \dot{\ell}_{i0} &= -\frac{1}{\tau_a}(\ell_{i0}-\lbar)-v \tanh\left(\frac{c_i -c_i^{0}} {c_i^{0} }\right) \;,\label{eq:l0}\\ 
    \dot c_i&=-\frac{1}{\tau_c}(c_i-c_i^{0})+\tilde\beta \tanh\left(\frac{\ell_i-\lbar}{\lbar}\right)\;.\label{eq:ci}
\end{align}
The rest length relaxes to $\bar\ell$ on time scales $\tau_a$ and is decreased at speed $v>0$ by myosin recruitment. In one dimension the rest length is related to the familiar active stress as
\begin{equation}
    \label{eq:sigma-a}
    \sigma_{ai}=-k(\ell_{i0}-\lbar)\;.
\end{equation}
On times long compared to $\tau_a$, Eq.~\eqref{eq:l0} then gives $\sigma_a\simeq \tilde\alpha\tanh[(c_i-c_i^{0})/c_i^{0}]$ with $\tilde\alpha=v k\tau_a$ the activity. It is then clear that $\tilde\alpha>0$ corresponds to an active contractile stress. Importantly, the relation between active stress and rest length only holds in one dimension. In higher dimensionality, stress is a tensor and needs to be defined more carefully and generally. 

The concentration $c_i$ of phosphorylated myosin turns over on the time scale $\tau_c$ to an ``equilibrium'' value $c_i^{0}$ that corresponds to  the unstressed system, with $\ell_i=\lbar$. Myosin couples to mechanical strain and is activated by cell elongation ($\ell_i>\lbar$) \rv{\cite{Mizutani2009RegulationCascade}} and degraded by cell contraction ($\ell_i<\lbar$) \rv{\cite{Takemoto2015CompressivePathway}} at  a rate per unit length $\tilde\beta>0$.

It is useful to also write the continuum equations corresponding to the above discrete model. To do this we nondimensionalize fluctuations in the myosin concentration as $\delta c_i=(c_i-c_i^{0})/c_i^{0}$. We also use $\lbar$ as unit of length, the relaxation time $\tau=\zeta/k$ that describes the rate at which mechanical stresses diffuse in the tissue as unit of time, and $k\lbar^2$ as unit of stress. The continuum equations can then be written in dimensionless form as
\begin{align}
    \label{eq:eppde}
   \partial_t \epsilon &= \partial_x^2 \epsilon + \partial_x^2 \sigma_a\;,\\
    \label{eq:eapde}
   \tau_a \partial_t \sigma_a &= -\sigma_a + \alpha \tanh(\delta c)\;,  \\
   \label{eq:c_pde}
    \tau_c \partial_t \delta c &= -\delta c + \beta \tanh(\epsilon)\;,
    \end{align}
where all quantities are now dimensionless.

The model now contains the two time scales $\tau_a$ and $\tau_c$ (measured in units of $\tau$) and two dimensionless parameters: $\alpha=vk\tau_a/\lbar$, which measures the ratio of the rate at which myosin activation builds up active stress to its decay rate,
and $\beta=\tilde\beta\tau_c/c^{0}$, which controls the rate of myosin activation due to mechanical elongation to the rate of intrinsic myosin turnover.
The model describes a generic activator-inhibitor system ($\delta c, \sigma_a$) coupled with a diffusive quantity $\epsilon$ that sets the scale of the chemical oscillations. On time scales much larger than the active time scales $\tau_a$ and $\tau_c$, Eq.~\eqref{eq:eppde} describes an overdamped solid with an effective elastic modulus $B_{eff}=k\lbar^2(1+\alpha\beta)$ enhanced by activity, as seen in experiments over very long timescales of 6-60 hours \cite{Vincent2015ActiveMonolayer}.

The model described by Eqs.~(\eqref{eq:eppde}-\eqref{eq:c_pde}) is related to ones previously studied in the literature~\cite{Boocock2020TheoryMonolayers,Armon2021, Banerjee2015PropagatingExpansion, Noll2017ActiveTissues, Munoz2013Physiology-basedViscoelasticity, Hino2020ERK-MediatedPolarization}. 
In particular, to linear order it is the same as the model introduced in  Ref.~\cite{Boocock2020TheoryMonolayers} to describe the emergence of spatiotemporal density waves in freely expanding monolayers from the activation of extracellular kinase activated by mechanical signals. Here we include saturating nonlinearities in both feedback terms in Eqs.~\eqref{eq:eapde} and \eqref{eq:c_pde}, which are important for stabilizing oscillations. 
 
There is some debate in literature about whether mechanical activation in cells is due to strain or strain rate feedback~\cite{Esfahani2021CharacterizationJunctions, Trepat2007UniversalCell, Kollmannsberger2011LinearCells, Vincent2015ActiveMonolayer, Borghi2012}.
Our model accounts for mechanical activation through both strain and strain rate. This can be seen by linearizing the above equations and formally integrating Eq.~\eqref{eq:c_pde}.  Substituting the result in Eq.~\eqref{eq:eapde} we obtain an integro-differential form of the dynamics of active stress, given by
\begin{align}
    \tau_a\partial_t \sigma_a &= -\sigma_a + \frac{\alpha\beta}{\tau_c} \int_0^t dt' e^{-(t-t')/\tau_c}\epsilon(t')\notag\\
    &\simeq -\sigma_a+\alpha\beta\epsilon-\alpha\beta\tau_c\dot\epsilon+\cdots \;,
    \label{eq:intdiff}
\end{align}
with $\dot\epsilon=\partial_t\epsilon$.
This form makes it evident that  active stress is controlled nonlocally in time by the entire mechanical strain history over the time $\tau_c$. \rv{Previous work on sustained oscillations incorporates an explicitly time-delayed strain coupling to the rest length~\cite{Munoz2018MechanicalChanges}. Equation \eqref{eq:intdiff} makes it evident that our model naturally incorporates this type of time delay.} We note that effective feedback through both strain and strain-rate is important for the emergence of sustained spatiotemporal waves (see Appendix A).

\rv{The numerical results reported below are obtained by 
solving the discrete nonlinear equations 
with RK4 timestepping and random initial conditions for the myosin levels of each individual cell.}
    
\begin{figure}
    \includegraphics[width=0.49\textwidth]{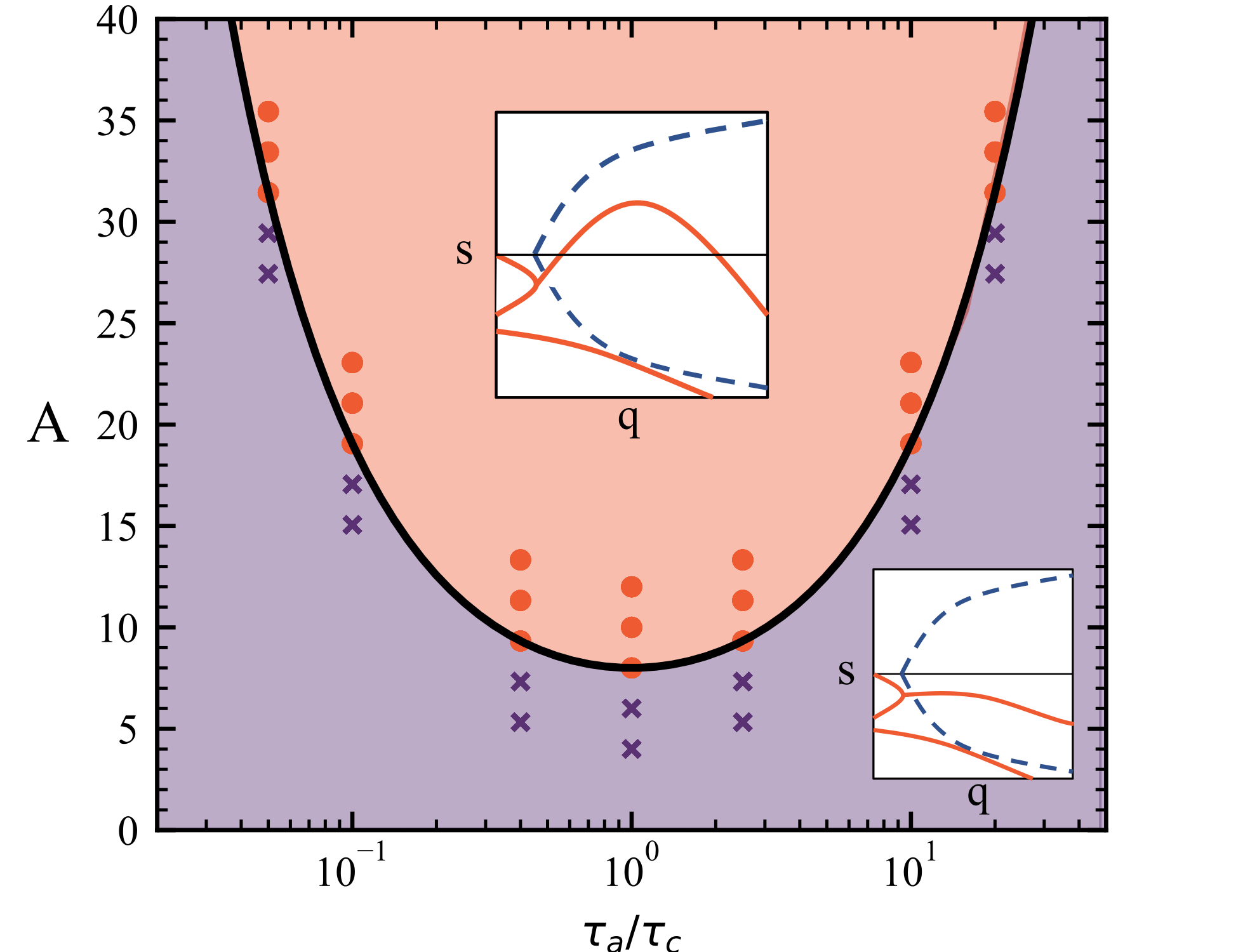}
    \caption{ 
   \rv{Phase diagram as a function of the active feedback $A=\alpha\beta$ and $\tau_a/\tau_c$ for a train of $N=100$ cells and
   $\tau=5$ min. The symbols are obtained via numerical solution of Eqs. (\eqref{eq:balance}, \eqref{eq:l0}, \eqref{eq:ci}), with  orange dots corresponding to sustained oscillations and blue crosses to diffusive decay. The solid line is the critical value $A_c$ obtained from linear stability analysis and given in Eq.~\eqref{eq:abcrit}. Insets: Real (solid orange lines) and imaginary (dashed purple lines) part  of the dispersion relations of the linear modes  for $A>A_c$ (top), where one the modes becomes unstable ($Re[s(q)]>0$) signaling the transition to
   sustained oscillations, 
   and $A<A_c$ (bottom) where all modes are stable. The linear instability for $A>A_c$ is stabilized by nonlinear terms, resulting in sustained oscillations.}}
    \label{fig:disp_phase}
\end{figure}

\section{Emergent Spontaneous Oscillations}
In the absence of external perturbations, the homogeneous steady state is simply $c=c^{0}$ and $\epsilon=\sigma_a=0$. Next we examine the linear stability of this base state by probing the linear dynamics of fluctuations in both unconfined and confined systems.

\subsection{Unconfined tissue}
We first consider an unconfined tissue and assume periodic boundary conditions. We linearize Eqs.~(\eqref{eq:eppde}-\eqref{eq:c_pde}) and look for solutions of the form $\epsilon,\sigma_a,\delta c\sim e^{s t+iqx}$.
The dispersion relations of the linear modes are then obtained from the solution of a cubic equation, given by
\begin{equation}
  ( s+q^2) (\tau_a s +1)(\tau_c s +1)=-q^2 \alpha\beta\;.
  \label{eq:disp}
\end{equation}
The onset of oscillations is controlled by the product of the parameters $\alpha$ and $\beta$ that govern the feedback between mechanical deformations and biochemical activation. We therefore refer to such a product $A\equiv\alpha\beta$ as active feedback. 

\rv{The dispersion relations obtained from Eq.~\eqref{eq:disp} are sketched in Fig. \ref{fig:disp_phase} (insets).} One of the modes is always real and stable due to the diffusive terms in Eq.~\eqref{eq:eppde}. We refer to this mode as $s_d(q)$. The other two modes can be imaginary, hence can show oscillatory behaviour. We refer to the pair of imaginary modes as $s_{1,2}(q)$ in the rest of the text.

The two  oscillatory modes, $s_{1,2}(q)$, become unstable for $A>A_c$, with 
\begin{align}
    \label{eq:abcrit}
    A_c &= \left(\sqrt{\frac{\tau_c}{\tau_a}}+\sqrt{\frac{\tau_a}{\tau_c}}\right)\left(\sqrt{\frac{\tau_c}{\tau_a}}+\sqrt{\frac{\tau_a}{\tau_c}}+2\right)\;,
\end{align}
 in a band of wavenumbers $q\in (q_-,q_+)$, given by
 \begin{align}
    \label{eq:qpm}
     q_\pm^2 &={} \dfrac{A-\dfrac{(\tau_a+\tau_c)^2}{\tau_a\tau_c}\pm \sqrt{(A-A_{c})(A-A_{c}+4\dfrac{\tau_a+\tau_c}{\sqrt{\tau_a\tau_c}})}}{2(\tau_c+\tau_a)} .
 \end{align}
The maximally unstable mode is the fastest growing mode, which we label as $s_{1,2}(q_c)=s_c, \bar s_c$. For $A>A_c$ the monolayer will spontaneously oscillate at the frequency of the fastest growing mode, given by $\omega_c=\text{Im}(s_c)$. The wavenumber and oscillation frequency of the fastest growing mode can be explicitly written for $A\sim A_c$ as
 \begin{align}
     \label{eq:qc}
     q_c^2 &=\frac{1}{\sqrt{\tau_a\tau_c}}\;, \\
      \label{eq:omegac}
     \omega_c^2 &= \frac{1 + q_c^2 (\tau_a +\tau_c)}{\tau_a \tau_c}~.
 \end{align}
 The parameter $A$ describes the ratio of active stress recruitment to passive stress dissipation. For $A>A_c$  active stress builds up faster than it is  dissipated by elastic diffusion, driving the instability. The instability requires mechanochemical feedback through both strain and strain rate. This can be achieved through the coupled dynamics of active stress and myosin, as  encoded by the coupled equations \eqref{eq:eapde} and \eqref{eq:c_pde},
 as done in Ref.~\cite{Boocock2020TheoryMonolayers}, or equivalently by an explicit coupling to strain rate in the dynamics of active stress, as shown in Eq.~\eqref{eq:intdiff}. The latter approach was effectively used in Ref.~\cite{Banerjee2019ContinuumMigration} by incorporating advection of actin-bound myosin by the network velocity. 
   
 This instability is stabilized by the saturating effect of the nonlinear couplings, resulting in the emergence of spontaneous sustained oscillations, which are obtained via numerical integration of the nonlinear equations. The value of $A_c$ obtained from linear stability analysis agrees well with the boundary between the region of decaying modes and sustained spontaneous oscillations obtained numerically, as shown in Fig.~\ref{fig:disp_phase}.

\begin{figure}[t]
    \includegraphics[width=0.5\textwidth]{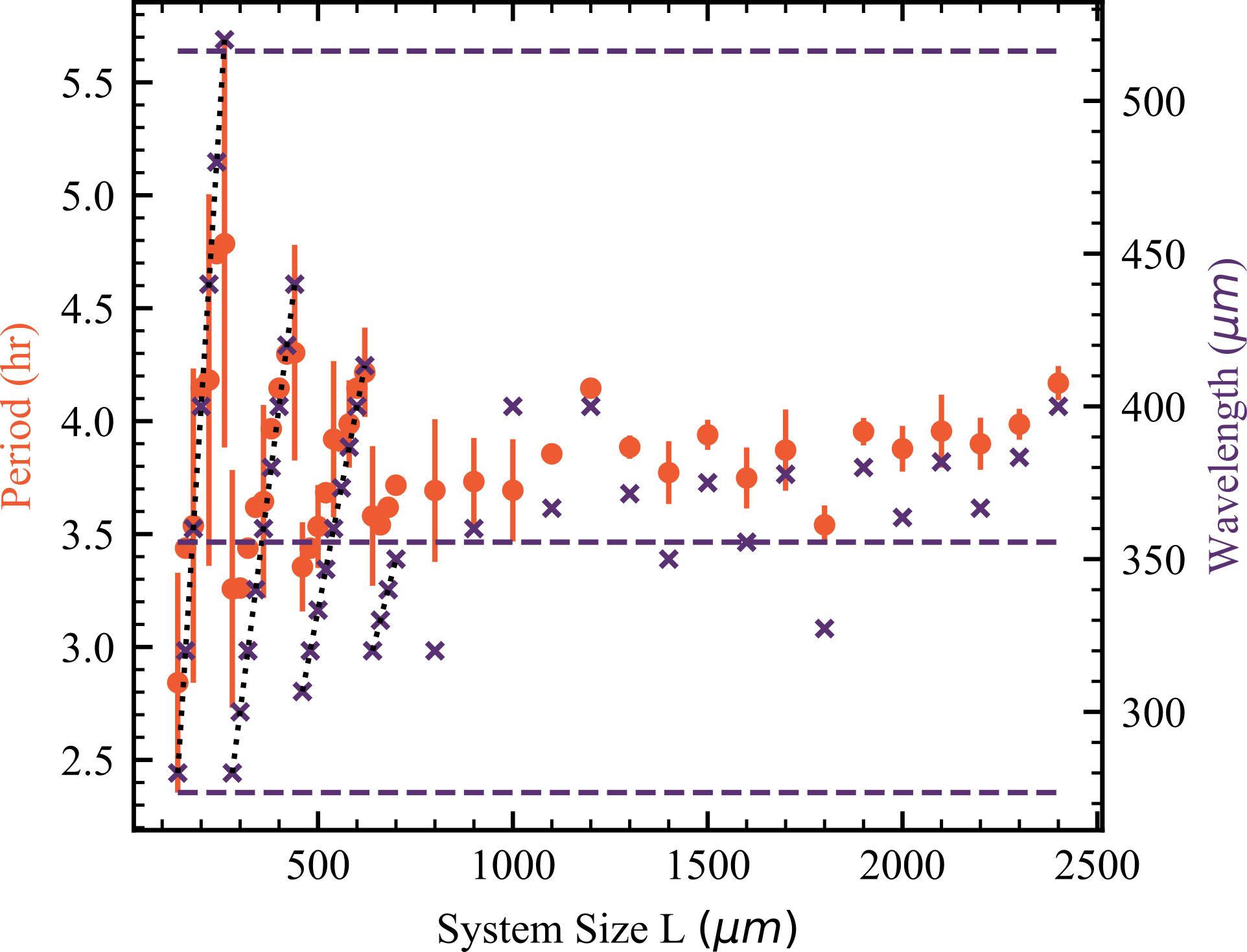}
    \caption{
    The length of a confined chain affects the nature of oscillations. The  period (orange dots, left axis) and wavelength (purple crosses, right axis) of the oscillations are plotted as functions of the system size $L$. The results are obtained by numerical integration of the nonlinear Eqs.~\eqref{eq:balance},~\eqref{eq:l0},~\eqref{eq:ci}.  The lower and upper dashed lines correspond to the wavelengths $\lambda_{min}$ and $\lambda_{max}$, respectively. The middle dashed line is the wavelength of the fastest growing linearly unstable mode. Parameters determined by Eq. \eqref{eq:qc},~\eqref{eq:omegac} to match wavelength and period from Ref~\cite{Petrolli2019Confinement-InducedModes}: $\tau=10 min$, $\tau_a=20 min$, $\tau_c=400min$, $A=33.4$, $\beta=1.0$.}
    \label{fig:harmonics}
\end{figure}

\subsection{Confined tissue}
For the confined system we use  Dirichlet boundary conditions for the displacement, $u(0,t)=u(L,t)=0$. In this case only wavelengths that are multiples of the system size $L$ are allowed, i.e., $\lambda_n=2L/n$, with $n$ an integer. Oscillatory unstable modes exists  for $A>A_c$ and $\lambda\in [\lambda_{min},\lambda_{max}]$, with $\lambda_{min,max}=2\pi/q_\pm$.  We therefore expect that for a fixed value of $L$, a confined 1-D cell train will support spontaneous standing waves corresponding to $n\in[n_{min},n_{max}]$, with $n_{min,max}=2L/\lambda_{max,min}$.  This is indeed shown in Fig.~\ref{fig:harmonics} that displays the period (orange dots) and wavelength (purple crosses) of oscillatory modes in a confined cell line as functions of the length $L$ of the system. The lower and upper dashed lines in the figure correspond to the values $\lambda_{min}$ and $\lambda_{max}$, respectively. The middle dashed line is the wavelength of the fastest growing mode $\lambda_c=2\pi/ q_c$. For the parameter values used in the figure, $\lambda_{min}= 273.6 \mu m$, $\lambda_{max}= 516.3 \mu m$, and $ \lambda_c = 355.6 \mu m$ and no oscillations are observed for 
$L< 140 \mu m$. As $L$ increases, we observe oscillations of wavelength corresponding to the first 4 modes $n=1,2,3,4$ shown by the dotted black lines in the figure.  Each mode can only be excited provided $n\lambda_{min}/2<L<n\lambda_{max}/2$. When $L$ exceeds this value for $n=1$, the system spontaneously excites the second mode and so on, until $L\sim 4(\lambda_{max}-\lambda_{min})$. Above this critical value, we see a transition where the wavelength and period of spontaneous oscillations becomes independent of the system size as the standing waves of wavelength controlled by the system size are replaced by ``intrinsic'' spontaneous oscillations at the natural wavelength $\lambda_c$ of the tissue. The ratio between wavelength and period remains constant and provides an estimate for the speed of oscillations. To understand the origins of this critical length, we note that the preferred wavelength of the system is given by the fastest growing wavevector, but it is not always attainable due to the constraint imposed by confinement. For large systems, the band of allowed modes, $[n_{min}, n_{max}]$ is also large, such that a wavelength close to the fastest growing wavelength can be attained. For smaller systems, in contrast, the range of allowed modes is limited by the system size. The crossover between the two is controlled by the width of the band of unstable wavevectors, which in turn is controlled by the activity coupling $A$.

This behavior has been observed experimentally in confined 1D epithelia by us \rv{ (see SI video 1 and 2) and previously others~\cite{Petrolli2019Confinement-InducedModes}} and in simulations of the Self-Propelled Voronoi model~\cite{Petrolli2019Confinement-InducedModes}, and was understood as arising from a tendency of cells to align their direction of polarization with their own motility~
\cite{Ron2020One-dimensionalPatterns, Petrolli2019Confinement-InducedModes, LoVecchio2024SpontaneousBoundaries}. 
Here we show that mechanochemical couplings alone can capture oscillations in confined epithelia and the transition from size-dependent standing waves to intrinsic spontaneous oscillations even in the absence of cell-polarity or ad hoc aligning mechanisms. Probing such a transition in confined tissue can provide a way of extracting quantitative information on the strength of such mechanochemical couplings.

\begin{widetext}

\begin{figure}[t]
    \includegraphics[width=\textwidth]{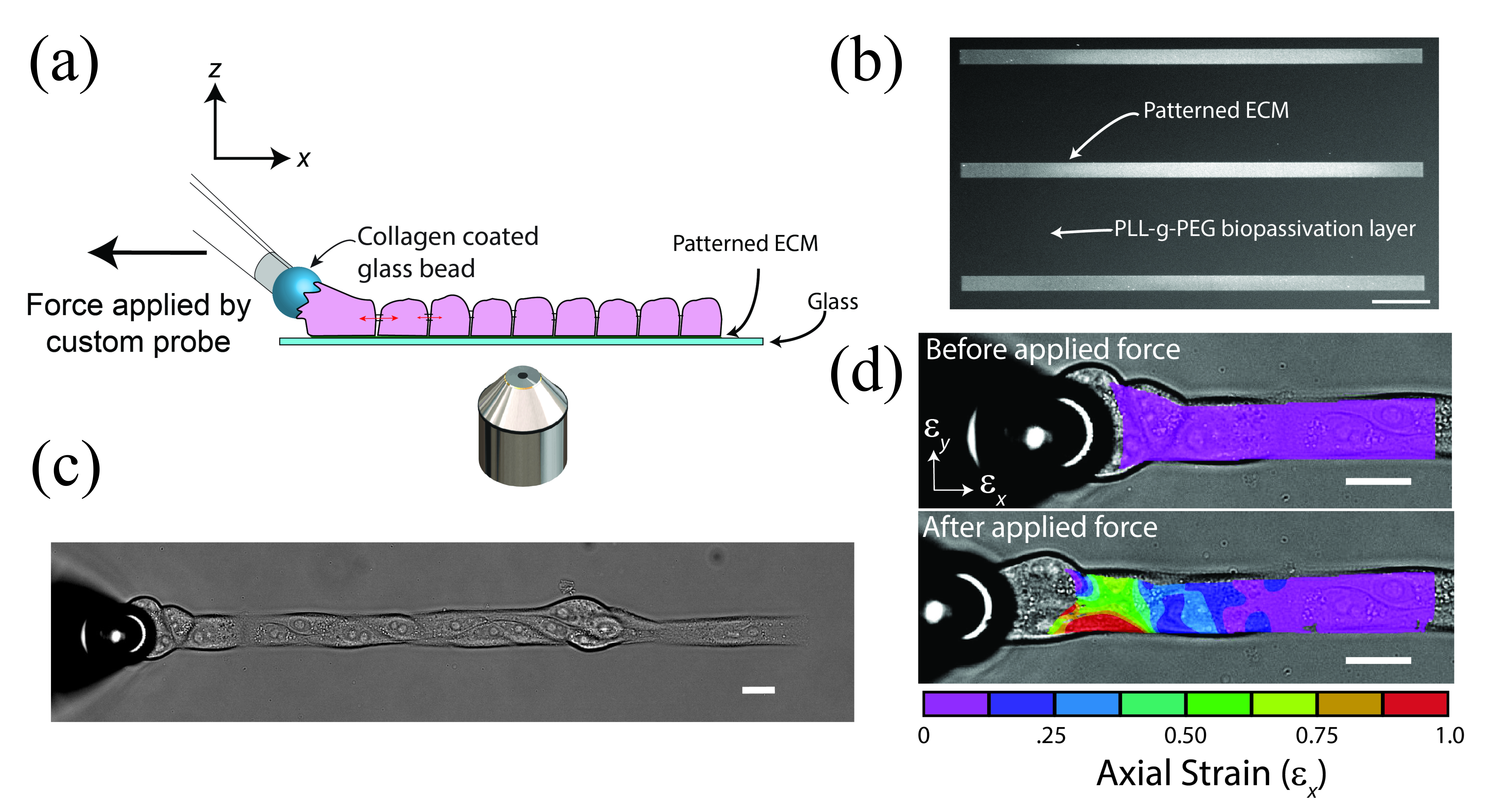}
    \caption{\rv{Micromanipulation of epithelial 1-D cell train. (a) Schematic of the 1-D model used for boundary force perturbation studies. (b) 390x20 $\mu m$ tracks for cells can be created in a high throughput manner using protein micropatterning, visualized using Oregon green gelatin. The ECM tracks constrain ~20 cells to a single file. (d) Following a $30\mu m$ displacement applied at a rate of at $1\mu m /sec$, only the outer 2-3 cells were strained, allowing us to investigate the effects of the transmission of boundary forces on the unstrained collective. Scale bars are (b) 50 $\mu$m and (c,d) 20 $\mu$m.}}
    \label{fig:step_strain_exptl}
\end{figure}
\end{widetext}

\section{Experimental Setup and Response to Step-Strain}
Next we examine the response of the system to externally imposed mechanical deformations.  
\rv{This is inspired by our recent experiments where  a 1-D micropatterned epithelial train of cells is stretched at one end  with an adhesive bead (see SI video 3 and 4)~\cite{ Dow2023UCContacts}.}
\rv{The custom experimental micromanipulation setup is shown in Fig. \ref{fig:step_strain_exptl}.  MCDK cells are confined to a 1-D geometry by an adhesive pattern of collagen I on a glass substrate. The probe is built by glueing a glass microbead to a pipette tip
and functionalizing the bead with collagen I (Fig. \ref{fig:step_strain_exptl}a). 
The probe is then mounted onto a piezo-driven micromanipulator. Once the probe forms a stable adhesion to one end of the MDCK cell train (Fig. \ref{fig:step_strain_exptl}c), the micromanipulator pulls the probe $30 \mu m$ (approximately 1.5 cell lengths) stretching the cell line. This strains the local cell junctions  at physiological rates, as quantified using digital image correlation DIC (\ref{fig:step_strain_exptl}d). This procedure generally exerts a direct strain only on the outer 2-3 cells (see SI video 3)(Fig.\ref{fig:step_strain_exptl}d). 
We patterned lines of 20x390 $\mu m$ to ensure lateral confinement of the cells (Fig. \ref{fig:step_strain_exptl}b). The length is short enough to investigate the effect of mechanical forces on the entire system.}

\begin{figure}[t]
    \includegraphics[width=0.48\textwidth]{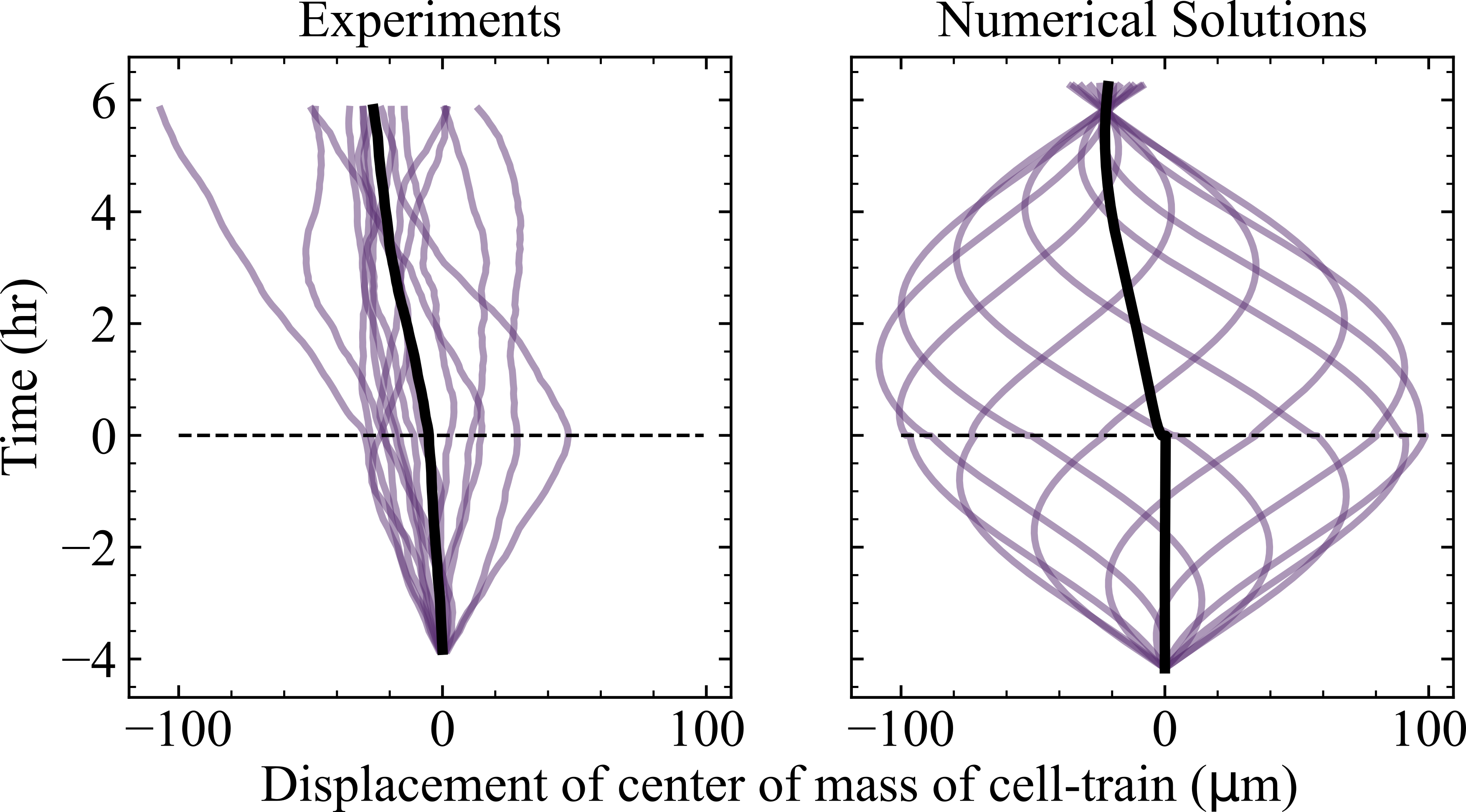}
    \caption{The cell-train's center of mass moves towards the direction of the applied step-stretch. \rv{Both plots show the instantaneous displacement of the center of mass of a 1-D cell train to an external stretch  applied at $t=0$ along the $-x$ direction. The light purple lines are the trajectories of individual experiments/simulations. The dark blue line is the average behavior. In both cases there is high variability in the response of individual cell trains,  as the cell train largely continues to move in the direction determined by the spontaneous oscillations, although with an enhanced amplitude when the latter are towards $-x$.  The mean motion, however, is always towards the externally applied step-stretch. Experiments: trajectories are obtained from PIV measurements averaged over $n=11$ separate experiments where the contact between the pulled cell and rest of the cell line did not dissociate. Simulations: the trajectory of the center of mass is averaged over $n=100$ simulations (only $n=11$ trajectories are shown).
    }
    The simulation parameters are chosen to match wavelength and period from experiments: $\tau = 5$ min, $\tau_a = 100$ min, $\tau_c = 250$ min, $A=10.5$, $\beta = 1.0$, $L=400 \mu$ m, magnitude of step strain = $30 \mu $m.}
    \label{fig:step_strain}
\end{figure}

 To quantify the position of the cells, we choose an $x$ axis oriented along the 1-D cell train with $x=0$ corresponding to the left edge of the unstretched cell train (Fig,~\ref{fig:step_strain_exptl}(a)). We then stretch the tissue by pulling towards $-x$ and choose $t=0$ as the time of application of the stretch. Using Particle Image Velocimetry (PIV), we tracked the movement of the cell train over the course of 6 hours following the localized stretch on the cell train. The local stretch resulted in the propagation of the strains through the tissue. \rv{In the 6 hours following the stretch, the entire cellular system moved approximately 23 $\mu$m in the direction of the applied stretch.}

We find that the response of individual cell-trains
depends on whether the perturbation is applied in-phase or out-of-phase relative to the spontaneous oscillations of the monolayer. If the stretch is applied while the system is spontaneously moving in the same direction as the applied stretch (i.e., towards $-x$), then the cells continue to move towards $-x$ before reversing due to spontaneous oscillations, although the period of oscillation is slightly increased by the applied 
stretch that increases the length of the cell train. If the stretch is applied while the boundary cells are moving in the direction opposite to the applied stretch (i.e., towards $+x$), they will continue to move in the same direction. \rv{In both cases we observe a small delayed  response  to the external stretch in the form of a slight increase of the amplitude of oscillations towards negative $x$. This is supported by the simulations (Fig.~\ref{fig:step_strain}). The fact that the position of the center of mass of the cell trains averaged over many replications in both experiments and simulations shows motion towards the step-stretch direction implies that the cell-trains are not entirely insensitive to the pull. The individual response, however, is highly variable and depends on the instantaneous phase of spontaneous oscillations relative to the step-stretch direction.}

It is evident that in both experiments and simulations, a step perturbation does not provide quantitative information of the activated processes that control the tissue dynamics because the strains and stresses induced in the tissue decay on time scales of the order of the passive stress relaxation $\tau$. The relaxation time $\tau$ is generally much shorter than the times scales controlling the build up of active stress and myosin turnover. In addition, in our experiment to avoid losing adhesion or ripping the tissue, the applied stretch must be smaller than the amplitude of spontaneous oscillations, making it difficult to quantify the effect of this perturbation on the tissue. \rv{These findings are  consistent with the recent observation that activation of leader cell polarity in expanding cell trains decays within a few cells and that this local action is not sufficient to  guide the entire cluster~\cite{Rossetti2024OptogeneticMigration}.} 

To probe the dynamics of the tissue on a broad range of time scales we need an applied perturbation capable of spanning this range of time scales. For this reason in the next section we examine the response to oscillatory boundary perturbations. 

\section{Response to Oscillatory Boundary Perturbations}
Next we examined the response of the system to an oscillatory strain applied at the boundary. For simplicity we consider a symmetric perturbation applied at both ends of a system of length $2L$. By symmetry the response is the same as for an oscillatory perturbation applied to a system of length $L$ while keeping the other end fixed.

To evaluate the linear response to oscillatory perturbation, it is convenient to consider equations for the displacement $u(x,t)$ instead of the strain. Linearizing Eqs.~(\eqref{eq:eppde}-\eqref{eq:c_pde}), and using the same dimensionless units, the dynamics is then governed by
\begin{align}
    \label{eq:upde}
   \partial_t u &= \partial_x^2 u + \partial_x \sigma_a\;,\\
    \label{eq:eapde_lin}
   \tau_a \partial_t \sigma_a &= -\sigma_a + \alpha \delta c\;,  \\
   \label{eq:c_pde_lin}
    \tau_c \partial_t \delta c &= -\delta c + \beta \partial_x u\;,
    \end{align}
To describe an oscillatory mechanical perturbation at the boundaries, we solve these equations with boundary conditions $u(0,t)=-u(2L,t)=\Delta \sin{(\omega_0 t)}$. We additionally require $\left[\partial_x\delta c(x,t)\right]_{x=0, 2L}=0$ and $\left[\partial_x\sigma_a (x,t)\right]_{x=0, 2L}=0$. These conditions guarantee no flux at the boundaries. The details are given in Appendix B.
We work in Fourier space and take the time Laplace transform with complex frequency $s$. Only even components of the amplitudes of the displacement, $\hat{u}(q_n,s)$, with $q_n=2n\pi/2L$, are nonzero and can be written in terms of a wavenumber and frequency dependent susceptibility as

\begin{align}
    \hat{u}(q_n,s) &={} \hat\chi(q_n,s) \hat{F}(q_n,s)\;,
        \label{eq:ufourier}
\end{align} 
where 
\begin{align}
    \hat{F}(q_n,s)= -\dfrac{2\Delta }{n\pi}\frac{s\omega_0}{\omega_0^2 + s^2}\;.
\end{align} 
The susceptibility $\hat\chi(q_n,s)$ is naturally written in terms of a frequency-dependent inverse penetration length $b(s)$ as
\begin{align}
    \hat\chi(q_n,s) &={} \dfrac{b^2(s)/ s}{b^2(s) + q_n^2}\;,
    \label{eq:chi}
\end{align}
with
\begin{align}
    b^2(s) &={} \dfrac{s}{1+\dfrac{A}{(1+s\tau_a)(1+s\tau_c)}}\;.
    \label{eq:b}
\end{align}

\begin{figure}[t]
    \centering
    \includegraphics[width=0.5\textwidth]{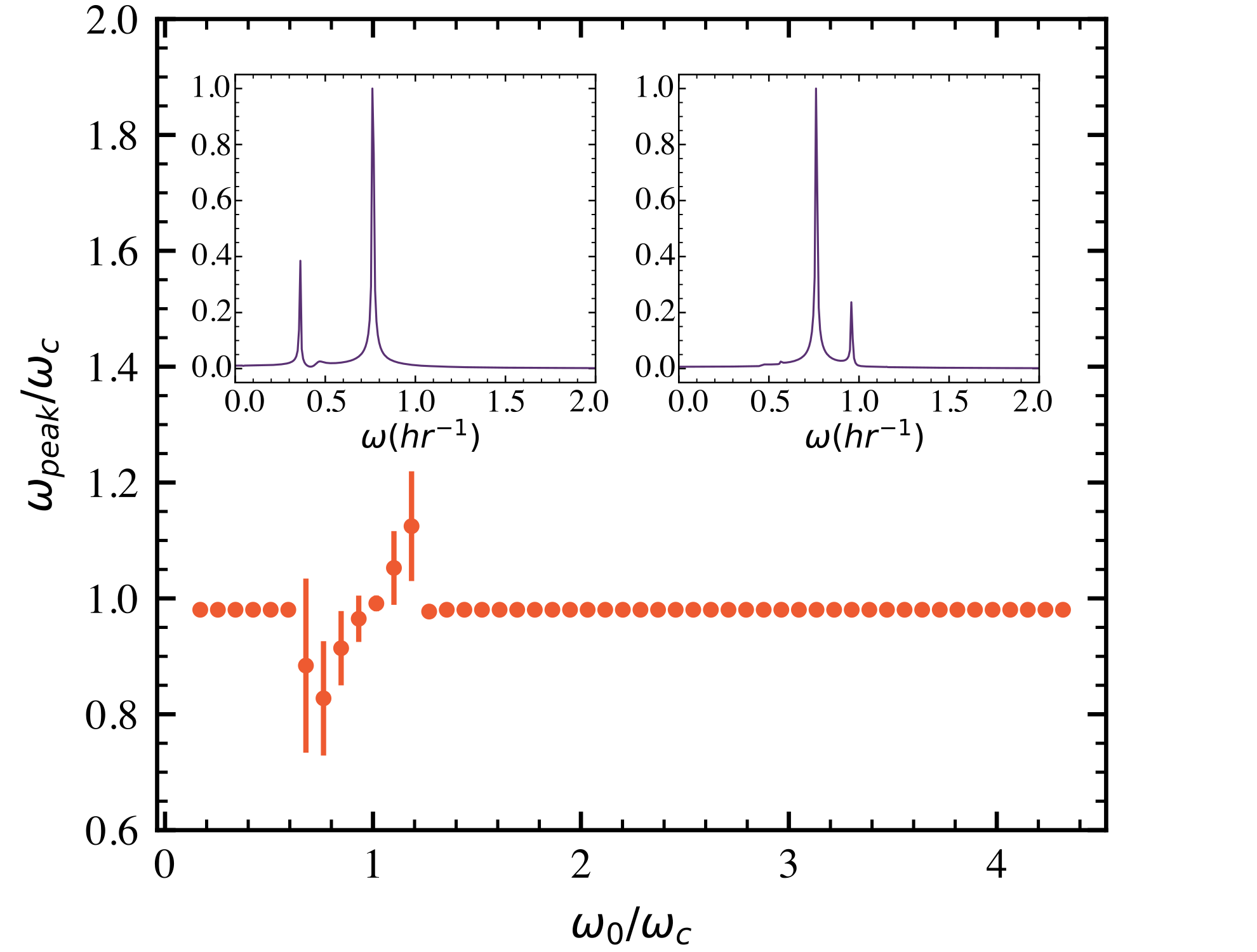}
    \caption{ The frequency $\omega_{peak}$ of the largest peak in the Fourier spectrum $\tilde{u}_i(\omega)$ of individual cell displacement is plotted as a function of the frequency $\omega_0$ of the external perturbation. Both frequencies are scaled by the frequency $\omega_c$ of the spontaneous oscillations of the tissue. The plot shows that $\omega_{peak}=\omega_c$ other than when $\omega_0\sim\omega_c$, where it becomes difficult to distinguish the two peaks.
    Insets: 
    The spectrum of the displacement field $\tilde{u}_i(\omega)$ shows clear peaks at $\omega=\omega_0$ (smaller peak) and $\omega=\omega_c$. Left inset: $\omega_0=0.36 hr^{-1}$. Right inset: $\omega_0=0.96 hr^{-1}$.   
    Parameters used: same as Fig.~( \ref{fig:step_strain}) which give  $\omega_c=0.67 hr^{-1}$, and $L=600\mu m$.}
    \label{fig:peaks_num}
\end{figure}

The real part of $b$  determines the penetration length $\ell_p(\omega)=[\text{Re}(b(i\omega))]^{-1}$ of the boundary perturbation.  In the absence of activity {$A=0$}, one recovers the response of an overdamped solid, with  a susceptibility $\hat\chi(q_n,s)=1/(s+q_n^2)$ describing simple diffusive dynamics and a penetration length  $\ell_p = \sqrt{\frac{2}{\omega}}$.
Recalling that frequency here is measured in units of the diffusive time $\tau$, this implies that  perturbations at frequency large compare to $\tau^{-1}$ are quickly damped, while low frequency perturbations travel throughout the solid. For a typical epithelial tissue, we estimate $\tau\sim 5 min$ \cite{Banerjee2015PropagatingExpansion}. 

It is instructive to invert the Fourier and Laplace transforms, to obtain the displacement as a function of space and time. To do this, we examine the behavior for $A$ just above $A_c$ and only include the contribution from the poles $s=\pm i\omega_0$ associated with the external perturbation and from the fastest growing natural excitations or modes of the unperturbed system, $s=s_c$. The details of the calculation are given in Appendix C, with the result

 \begin{widetext}
\begin{align}
    \label{eq:uoscsol}
    u(x,t) ={}& \Delta|\mathcal{B}(x)|\sin{(\omega_0t + \phi_1)} 
    - 4\Delta\frac{ \sin{(m_c\pi x/L)}}{m_c\pi} \frac{\omega_0 e^{\text{Re} (s_c) t}}{|(s_c^2+\omega_0^2)\mathcal{T}(s_c)|} \cos{(\omega_c t - \phi_2)} + \text{transient modes}\;,
\end{align}

where $\mathcal{B}(x)=|\mathcal{B}(x)|e^{i\phi_1}$ is a complex amplitude that encodes the spatial dependence of the response and $\mathcal{T}(s)$ is a complex time scale. Their explicit expressions, together with the definition of the phase $\phi_2$, are given in Appendix B. The integer $m_c$ is defined as  $ m_c = \lfloor q_c L/\pi \rfloor$, where $\lfloor x\rfloor$ represents the floor of $x$, and $q_c = (\tau_a \tau_c)^{-1/4} $ is the wavenumber of the fastest growing mode at $A=A_c$. The frequency of oscillation of the second term $\omega_c$ is the spontaneous oscillation frequency from the previous sections (see Appendix C).
\end{widetext}

The response to the perturbation naturally separates into two contributions. 
The first term on the right hand side of Eq.~\eqref{eq:uoscsol} arises from the poles $s=\pm i\omega_0$ associated with the external periodic forcing, $\hat{F}(q_n, s)$. It describes a contribution that oscillates at the frequency $\omega_0$ of the external perturbation. We will refer to this term as the perturbation mode.  The second term originates from the poles of the susceptibility, $\hat\chi(q_n, s)$. It oscillates at the frequency $\omega_c$ of the spontaneous oscillations of the unperturbed system.  It will be referred to as
 the natural mode.
Of course the susceptibility has an infinite number of poles, but near $A_c$ an external perturbation mainly excites oscillations at the frequency of the external perturbation and at the frequency of the natural oscillations of the tissue. All other modes are rapidly damped.

This finding is supported by the numerical results shown in Fig.~\ref{fig:peaks_num}. The numerical solution is obtained by solving the full nonlinear system of Eqs.~\eqref{eq:balance},~\eqref{eq:l0}, ~\eqref{eq:ci} for a chain of length $L$, with $u(L,t)=0$ and $u(0,t)=\Delta\sin\omega_0 t$.  
We compute the displacement of each cell vertex, $u_i(t)$, at long times after the initial transients have died out. 
The Fourier spectrum of individual displacements, 
$\Tilde{u}_i(\omega)$, shown in the insets of Fig. \ref{fig:peaks_num} for a representative value of $i$ and two different $\omega_0$s displays two clear peaks at the external frequency $\omega_0$ (smaller peak) and at the natural frequency $\omega_c$. 
Of course the two peaks merge when $\omega_0\sim\omega_c$. This is highlighted in the main part of Fig.~\ref{fig:peaks_num} that shows the frequency of the largest peak only deviates from $\omega_c$ when $\omega_0\sim\omega_c$ and it is not possible to distinguish the two peaks.

\begin{figure}[t]
    \centering
    \includegraphics[width=0.5\textwidth]{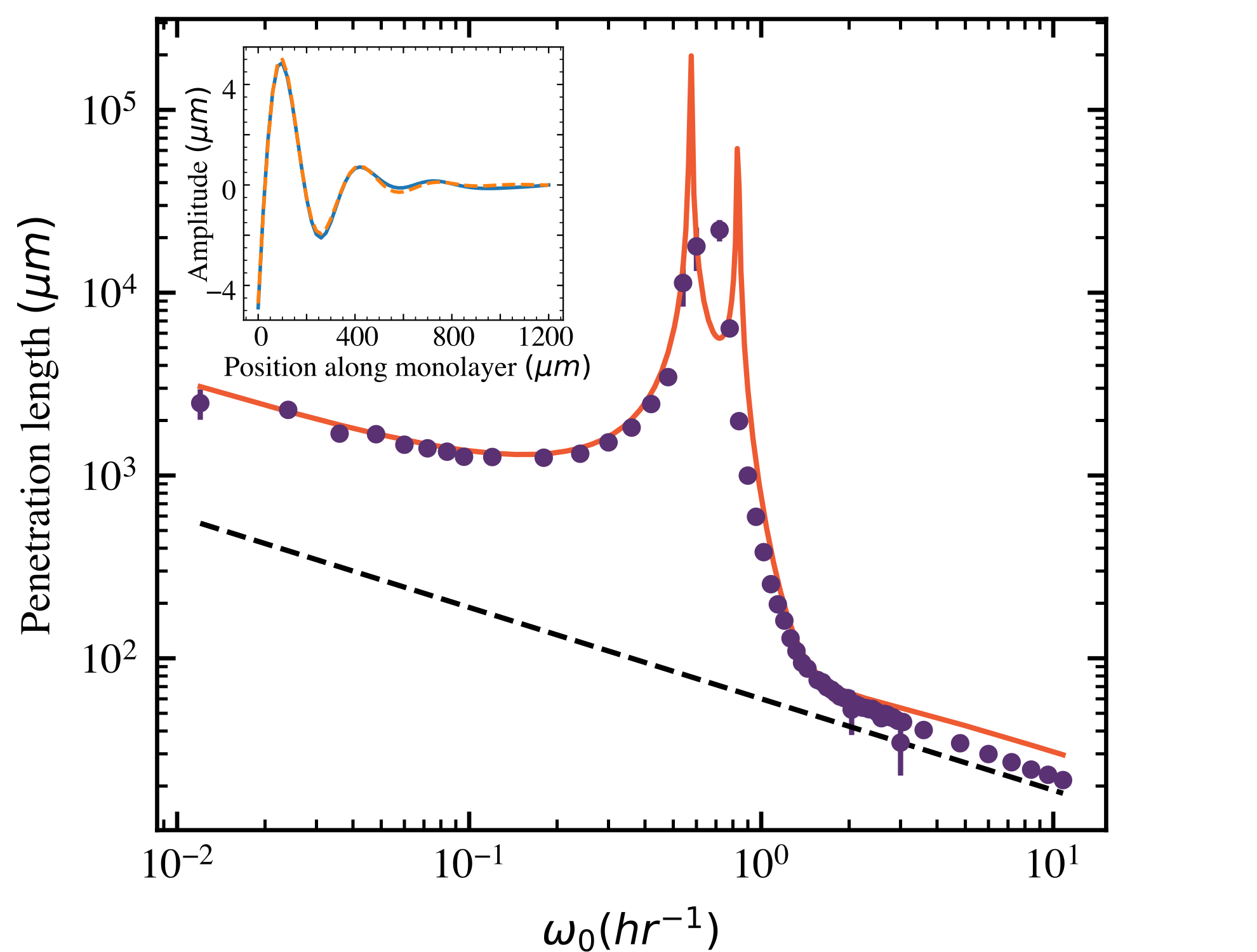}
    \caption{
   The penetration length extracted from numerics (purple dots) agrees well with the prediction of the linear theory (orange line) and decreases as $1/\sqrt{\omega_0}$  (black dashed line) at both low and high frequency. 
   Inset: Plot of the spatial profile of the perturbation mode of displacement field. The dashed line is a fit to an exponentially decaying sine function. Parameters used: same as Fig.~(\ref{fig:step_strain}) except $\Delta = 20 \mu m$, L=6mm, 8mm. Inset: L=1.2mm.}
    \label{fig:p_len}
\end{figure}

To quantify the effect of the external perturbation, we extract the contribution at $\omega_0$ from all numerically calculated displacements and examine its spatial dependence. 

According to our linear theory, the spatial decay of the first term of Eq.~\eqref{eq:uoscsol} described by $\mathcal{B}(x)$ is well fit by an exponentially decaying sine function, $\sim e^{-x/\ell_p}\sin(cx+d)$, for $\ell_p \gtrsim L$, with $c$ and $d$ fitting parameters. 

Fitting the numerical solution to this form, we compute the penetration length shown as blue dots in Fig.~\ref{fig:p_len}. The solid red line is the analytical expression obtained from linear theory. This diverges at the two values of frequency where the imaginary part of the linear modes vanishes. This corresponds to the vanishing of the imaginary part of the susceptibility that describes dissipation, where the perturbation is free to travel undamped across the entire system.

We can only reliably find the penetration length when  $\omega_0$ is well separated from the natural oscillation frequency, since otherwise the two modes overlap. 
At small frequencies, that is on time scales slower than any of the time scales of the system, active stress and myosin equilibrate quickly and one can write $\sigma_a\simeq A \epsilon_p$.  This simplifies the strain dynamics and gives $\ell_p \simeq \sqrt{(1+A)/\omega}$. In this regime the system responds like an overdamped solid, but with a  stiffness of the system enhanced by contractile activity. 
At large frequencies the active stress and myosin do not have time to respond and give $ \ell_p\simeq \sqrt{1/\omega}$, as in a passive overdamped solid. 
The difference in intercepts of the left and right linear-sections can provide a direct measure of the active mechanochemical coupling, $A$, in physical systems.
 The penetration length of an oscillatory boundary perturbation can therefore provide a direct probe of the tissue dynamics.
 On the other hand, one would need to probe the system at very low frequencies ($\sim 10^{-1}-10^1 hr^{-1}$), 
 which may be challenging.
  A square wave with the same frequency and time period as the sine wave perturbation may provide a more realistic implementation and allow one to extract the same information, as shown in Appendix F. 

\begin{figure}[t]
    \centering
    \includegraphics[width=0.5\textwidth]{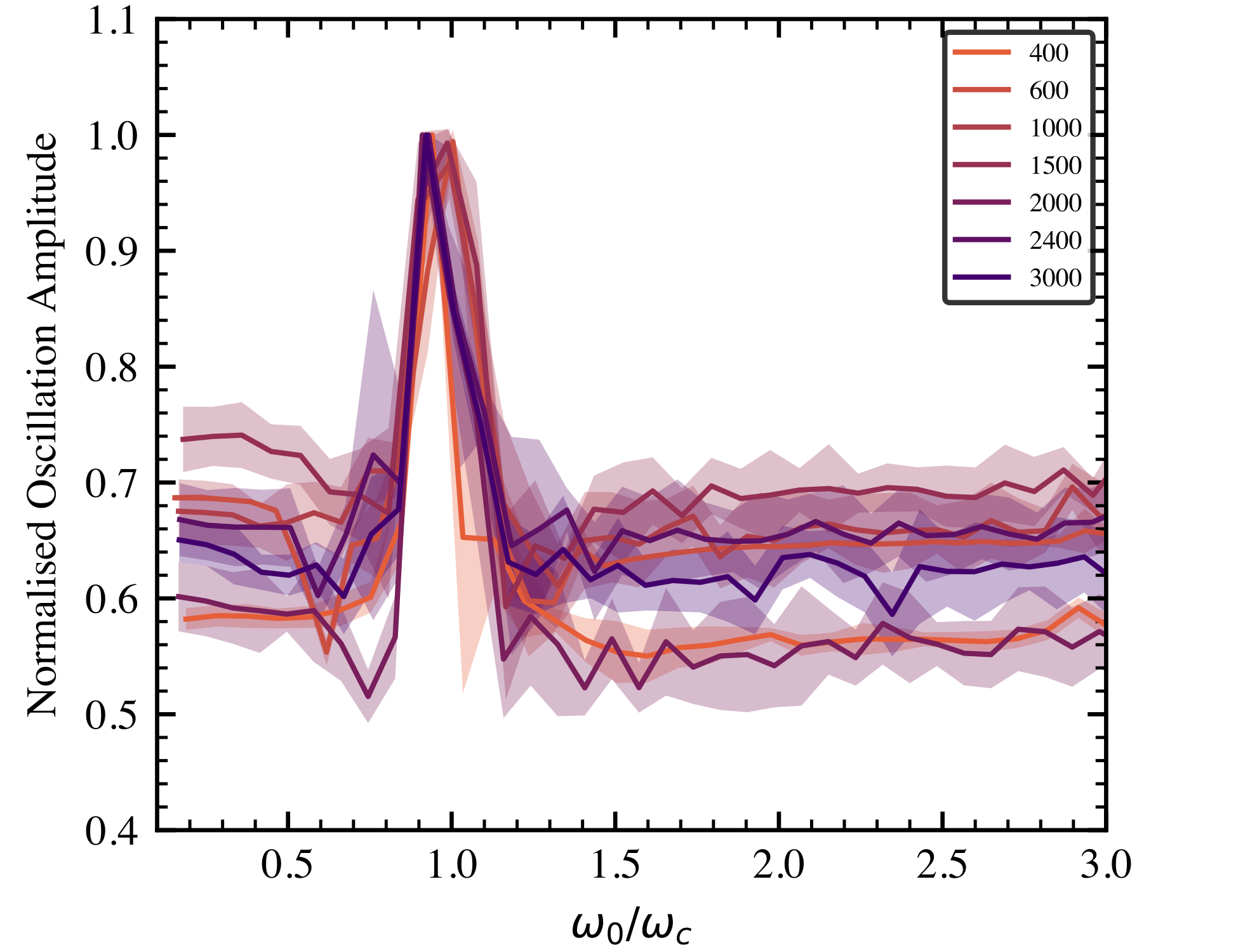}
    \caption{The amplitude of the response at any point along the cell train shows the resonance at natural frequency $\omega_c$. The various curves corresponds to different  cell-trains of lengths, L ($\mu m$). The shaded regions represent standard deviations. Parameters used: same as Fig.~(\ref{fig:step_strain}) but with $\Delta=L/40$.}
    \label{fig:resonance_num}
\end{figure}

We next examine the amplitude of the contribution at the natural frequency $\omega_c$. In our linear theory, for $A=A_c$, $\text{Re}(s_c)<<\text{Im}(s_c)$, thus the amplitude of the natural mode becomes inversely proportional to $(\omega_0^2-\omega_c^2)$, showing a resonance when the perturbation frequency $\omega_0$ matches the natural frequency $\omega_c$. This resonance is reproduced in the numerics, as shown in Fig.~(\ref{fig:resonance_num}). Here we plot the amplitude of the contribution from the natural mode to the response at a fixed point in space as a function of $\omega_0$ for several system sizes. The highest amplitude for each line is scaled to 1. The response shows a clear peak at $\omega_c$. Since the numerical system is nonlinear, we also see phase locking at resonance, where the value of the locked phase varies with the length of the system.

\section{Concluding Remarks}

We have studied a minimal model of 1-D cell trains that exhibit spontaneous and sustained spatiotemporal waves, and we confirm these oscillations and responses to perturbations are observed in our experimental system of a confined 1-D cell train of MDCK epithelial cells~\cite{Dow2023UCContacts}. We further examine both analytically and numerically the response of this model to an external sinusoidal boundary perturbation.
We find that when driven by a periodic boundary perturbation of frequency $\omega_0$, the tissue exhibits resonance at its natural frequency, $\omega_c$, which is controlled by the interplay of time scales associated with myosin recruitment and activation. Oscillatory boundary perturbations can therefore provide a method to infer such time scales.
We additionally show that the penetration length of the perturbation can provide an important measurable quantity and allows an estimate of the mechanochemical couplings that control the behavior of the tissue.
In numerical simulations we also observe several phenomena typical of driven nonlinear oscillators, such as weak resonance at multiples of the natural frequency and phase-locking at resonance. The frequencies at which the effects of mechanochemical coupling are evident in Fig.~ \ref{fig:p_len} are on the time scale of active processes, which are very slow and estimated to be on the order of $10^{-1}-10^{1} hr^{-1}$. This may make it difficult to probe the response to smooth sinusoidal perturbations. The response could, however be probed via a square-wave perturbation, which should tease out all the effects described (see Appendix F).

\section{Acknowledgements}

We acknowledge support from The Institute of Collaborative Biotechnologies via grant ICB-2022-BEM-13. 

\appendix

\section{Linear stability analysis}

To evaluate the dispersion relations of the linear modes, we need to solve the cubic eigenvalue equation \eqref{eq:disp}. This is done numerically. To gain analytical insight we note that above a critical value of $A$, the real branch of the dispersion relation crosses the $x$ axis for a band of wave-vectors $q_\pm$. By solving $Re[s(q_\pm) ]= 0$ together with the dispersion relation we obtain the expression for $q_\pm$ given in Eq.~\eqref{eq:qpm}. The requirement that $q_\pm$ be real and positive gives us the condition $A>A_c$ for the existence of oscillations. This can also be seen from the discriminant in Eq. \eqref{eq:qpm}.

To understand why we need a model with three dynamical quantities, where active stress couples to strain rate, we now discuss a system of two coupled equations for strain and active stress, with both strain and strain rate coupling, as shown in Eq. \eqref{eq:intdiff}. For clarity, the corresponding equations with the same nondimensionalization as in the main text, are given by

\begin{align}
    \partial_t \epsilon &= \partial_x^2 \epsilon + \partial_x^2 \ea\;, \\
   \tau_a\partial_t\ea &= -\ea + A\epsilon - A\tau_c \partial_t \epsilon \;.
\end{align}
We can write a second order PDE for the strain by eliminating the active stress
\begin{align}
    \tau_a \partial_t^2 \epsilon &+ \left[1-(\tau_a-A\tau_c)\partial_x^2\right] \partial_t \epsilon -  (1+A)\partial_x^2\epsilon=0\;.
\end{align}
The first term with the second time derivative of the strain represents an inertial force with an effective mass $\tau_a$ and acts as a memory of previous deformations. The dynamics can be interpreted as that of a damped oscillator with  local effective frictional dissipation $\eta(q) = 1+(\tau_a-A\tau_c) q^2 $ and an effective elastic stiffness is renormalized by activity,  $K(q) = (1+A)q^2$.
The dispersion relations of the corresponding eigenmodes are given by,
\begin{align}
    s = \frac{-\eta(q) \pm \sqrt{[\eta(q)]^2-4\tau_aK(q)}}{2\tau_a}
\end{align}
where $\eta(q)$ can change signs but $K(q)$ is always positive. The modes can become  oscillatory and unstable when $\eta(q)<0$ and the discriminant is negative. It can be easily shown that these two conditions are satisfied in a range of wavenumbers. Thus oscillatory instabilities, which are necessary for engendering sustained oscillations upon stabilization by nonlinearities, will be present in this system for $A>\tau_a/\tau_c$. Without a strain rate coupling, the effective friction becomes $\eta=1+\tau_aq^2$ and is always positive. Thus all oscillatory solutions are stable and decay. So the presence of an oscillatory instability requires a strain-rate coupling. Finally, we point out that the above two coupled equations become unphysical for negative viscosities, since the smallest scales ($q\to \infty$) become unstable. This issue is circumvented by using three coupled equations as we have done in the main text.

\rv{To highlight the  importance of feedback through strain coupling, we consider the following equations
\begin{align}
    \partial_t \epsilon &= \partial_x^2 \epsilon + \partial_x^2 \ea\;, \\
    \tau_a\partial_t\ea &= -\ea + \alpha \delta c\;,\\
    \tau_c\partial_t\delta c &= -\delta c + \beta \epsilon - \gamma \partial_t \epsilon\;.
    \label{eq:c-App}
\end{align}
The coupling of myosin concentration to strain rate in Eq.~\eqref{eq:c-App} seeks to dilute the pool of myosin upon stretching and concentrate it upon compression. The dispersion relation is again a cubic equation in $q$ given by \begin{align}
    (\tau_c s+1)(\tau_a s+1)(s+q^2) = -q^2\alpha(\beta-\gamma s)\;.
\end{align}
We can now perform the same analysis as in the main text where we use $\text{Re}(s(q_\pm))=0$ to solve for a pair of wavevectors. We then assert that these wavevectors must be real and positive. The resulting condition for sustained oscillations is again $A>A_c(\gamma)$, with,
\begin{align}
    A_c(\gamma) &= \sqrt{E_\gamma} (\sqrt{E_\gamma} + 2)~, \\ 
    E_\gamma &= \left[ \sqrt{\frac{\tau_a}{\tau_c}} + \sqrt{\frac{\tau_c}{\tau_a}} \right]\left[ \sqrt{\frac{\tau_a}{\tau_c}} + \sqrt{\frac{\tau_c}{\tau_a}} -\frac{\alpha\gamma}{\sqrt{\tau_a\tau_c}} \right]\;.
\end{align}
Here  $\gamma$ shifts the range of instability towards lower values of activity $A$, but does not qualitatively change the results. For this reason we have used  $\gamma=0$ in the main text. On the other hand, if we consider a system with only strain rate coupling  ($\beta=0$, $\gamma\not=0$), then the condition for sustained oscillations becomes $ \alpha\gamma > \tau_a  + \tau_c$. This model also gives sustained oscillations in a range of parameters, but in this case the modes remain unstable at the shortest wavelengths ($s(q\to\infty)>0$), which is clearly unphysical. In other words, the  coupling to strain ($\beta\not=0$) has the important function of stabilizing the system at small scales.  
}

\begin{figure}[h]
    \includegraphics[width=\columnwidth]{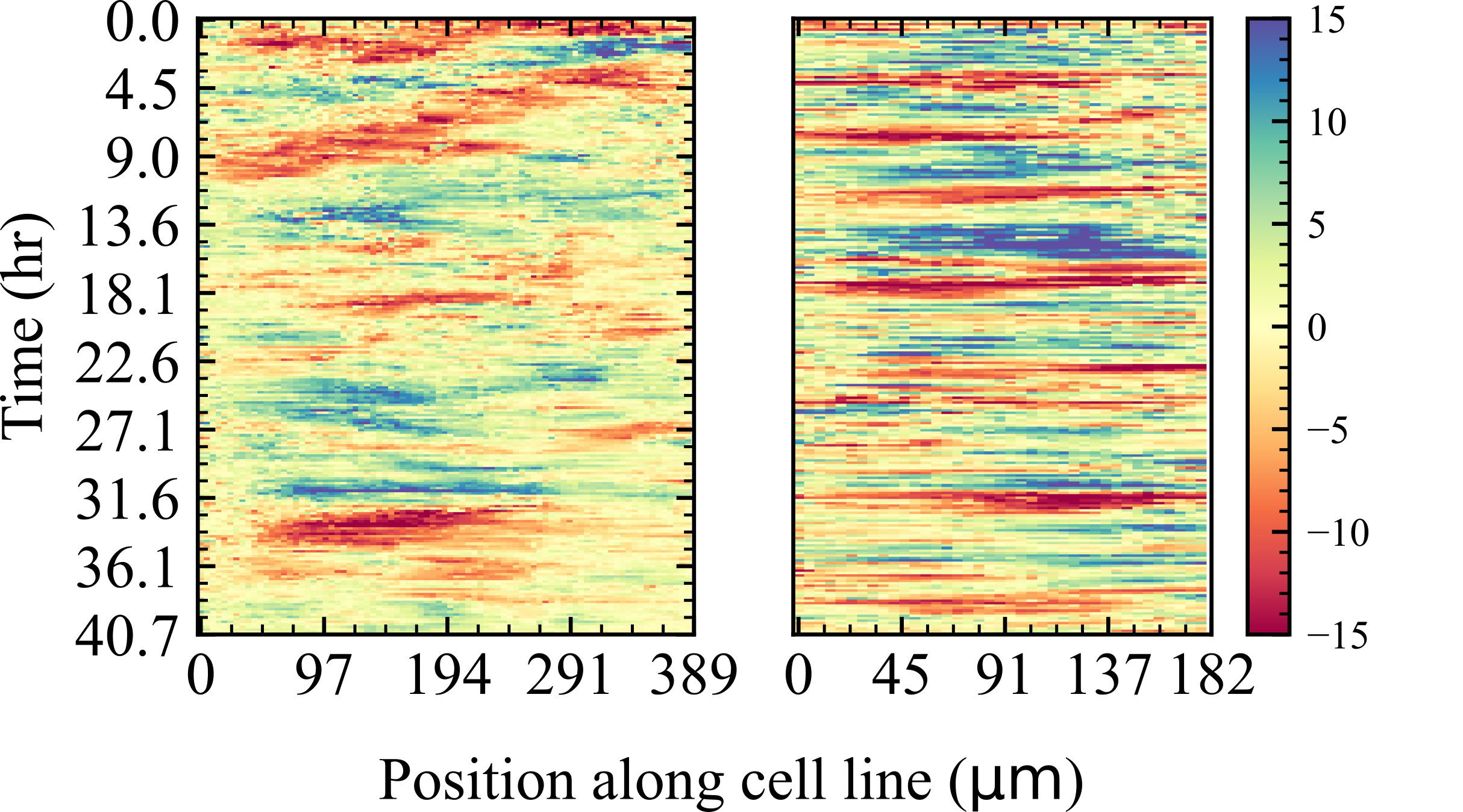}
   \caption{\rv{Kymographs of cell displacements obtained from PIV show global spontaneous oscillations for cell trains of length $390 \mu m$ (left) and $200 \mu m$ (right). In both cases the wavelength  is twice the system size. The period of the shorter cell train (right) is noticeably smaller than that of the longer one (left).}}
    \label{fig:SI_Kymographs_Standing_Wave}
\end{figure}

\begin{figure}[h]
    \includegraphics[width=\columnwidth]{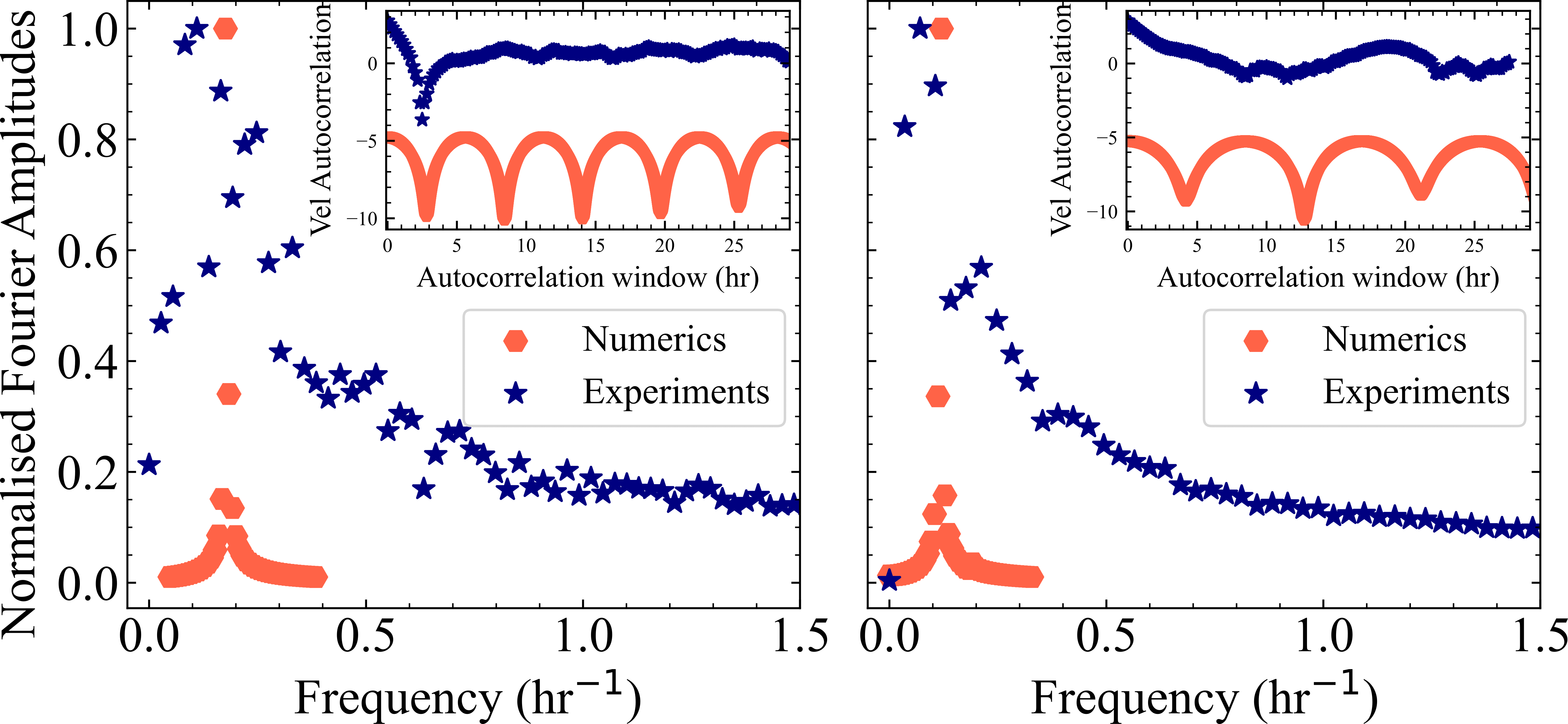}
    \caption{\rv{The Fourier decomposition of the velocity autocorrelation function of unperturbed cell trains shows clear peaks at the natural oscillation frequency.  Data are shown for two cell train lengths ($200 \mu m$ (left) and $390 \mu m$ (right)) from simulations (orange dots, n=10 runs) and experiments (navy stars, n=17 experiments). The experimental values peak around $T = 9.08$ hrs (left) and $14.2$ hrs (right) respectively. Inset: Plot of scaled logarithm of velocity autocorrelation function (blue curves are experimental, orange curves are numerical). The v-shaped periodic valleys are a signature of oscillations. Parameters used: same as Fig.\ref{fig:step_strain}.}}
    \label{fig:SI_FFT}
\end{figure}

\rv{
\section{Spontaneous Oscillations in Experiments}

Spontaneous oscillations of epithelia on 1-D micropatterned lines have been reported before \cite{Petrolli2019Confinement-InducedModes}. In this Appendix we display data showing that both our experimental system and our simulated model indeed exhibits spontaneous oscillations 
when unperturbed.
Figure \ref{fig:SI_Kymographs_Standing_Wave} shows experimental kymographs of cell displacement versus time for two cell-train lengths, $390 \mu m$ and $200 \mu m$. The kymographs show global oscillations corresponding to the first spatial Fourier mode, hence with wavelength twice the system size. Since our experimental data are noisy, we use established autocorrelation methods to extract the periodicity of the dynamics. To find the period of oscillations we calculate the time series of the velocity of the center of the mass of each cell train and its velocity autocorrelation function. In both experiments and simulations, the Fourier decomposition of this function (Fig. \ref{fig:SI_FFT}) shows a peak at the natural oscillation frequency of the system.  We find that the period of oscillations for the longer cell train is approximately twice that of the smaller cell train, as predicted by theory.
}


\section{Derivation of solution to an oscillating boundary perturbation}

For generality, we present here the solution of Eqs.~(\eqref{eq:upde}-\eqref{eq:c_pde_lin}) in response to general time-dependent forces applied at the boundaries of the system, i.e., for $u(0,t)=f_1(t)$ and $u(2L,t)=f_2(t)$ . At the end of this Appendix we will then specialize to the specific oscillatory and symmetric perturbation discussed in the main text. The other boundary conditions impose no flux and are as described in the main text.

To proceed, it is useful to introduce a change of variables, 
\begin{align}
    u(x,t)=v(x,t) + f_1(t) + \left[f_2(t)-f_1(t)\right]x/2L\;. 
\end{align}
The equations then take the form
\begin{align}
    & \partial_t v - \partial_x^2 v - \partial_x \ea =  G(x, t)\;,
     \\
    & \tau_a \partial_t \ea + \ea - \alpha \delta c = 0\;, \\
    & \tau_c \partial_t \delta c + \delta c - \beta \partial_x v = \frac{\beta}{2L} \left[f_2(t)-f_1(t)\right]\;,
\end{align}
with
\begin{align}
    G(x,t)=\left(\dot f_1(t)-\dot f_2(t)\right)x/2L - \dot f_1(t) 
        \label{eq:Gxt}
\end{align}
and boundary conditions
\begin{align}
    & v(0, t) = v(2L,t) = 0\;,\\
    & v(x, 0)=\left[f_1(0)-f_2(0)\right]x/2L - f_1(0)\;,\\
    &\left[\partial_x \delta c(x, t)\right]_{x=0} = \left[\partial_x \delta c(x, t)\right]_{x=2L} = 0\;,\\
    &\left[\partial_x \sigma_a(x, t)\right]_{x=0} = \left[\partial_x \sigma_a(x, t)\right]_{x=2L} = 0\;,\\
    &\delta c(x,0)=0\;,~~~~\sigma_a(x,0)=0\;.
\end{align}
The source term $G(x,t)$ in the equation for the shifted displacement $v$ is equivalent to an externally applied force.

Keeping the boundary conditions in mind, we expand our variables in a discrete Fourier series in space, 
\begin{align}
 &v(x,t) = \sum_{n=1}^{\infty} v_n(t) \sin{(q_nx)}\;,\\
 &\delta c(x,t) = \delta c_0(t)+\sum_{n=1}^{\infty} \delta c_n(t) \cos{(q_nx)}\;,\\
 &\ea(x,t) = \sigma_{an}(t)+\sum_{n=1}^{\infty} \sigma_{an}(t) \cos{(q_nx)} \;,
\end{align}
with $q_n=n\pi/2L$.
The equations for the Fourier amplitudes are then given by
\begin{align}
    \label{eq:Gnt}
        &\partial_t v_n(t) +  q_n^2 v_n(t) + q_n \sigma_{an}(t)= G(n,t)\\
    &\tau_a \partial_t \sigma_{an}(t) + \sigma_{an}(t) - \alpha \delta c_n(t) = 0~, \\
    &\tau_c \partial_t \delta c_n(t) + \delta c_n(t) - \beta q_n v_n(t) = 0~,  
\end{align}
with \begin{align}
    G(n,t)=\left[-\dot f_1(t) +\dot f_2(t) (-1)^n\right]/q_nL
\end{align}
and initial conditions
\begin{align}
    &v_n(0) = \left[-f_1(0)+f_2(0) (-1)^n\right]/q_nL,\\
    &\sigma_{an}(0)=0,~~~ \delta c_n(0)=0 ~\forall n \in \mathbb{N}~.
\end{align}

Our equations are reduced to a set of ODE that we solve introducing the Laplace transform of the Fourier amplitudes, defined as
\begin{align}
   \hat{v}_n(s) = \int_{0}^{\infty} dt~v_n(t) e^{-st}\;,~~v_n(t) = \int_{\delta - i\infty}^{\delta+i\infty} \frac{ds}{2\pi i} \hat{v}_n(s) e^{st}
\end{align} 
where $\delta>0$ is 
chosen to the right of all poles of $\hat{v}_n(s)$. 
We can solve for $\hat{v}_n(s)$, with the result
\begin{gather}
    \hat{v}_n(s) = \frac{b^2(s)}{\tau s} \frac{ F(q_n,s)}{b^2(s) + q_n^2}\;,
\end{gather}
where
\begin{align}
    b^2(s) = \frac{s}{1+\frac{A}{(1+s\tau_a)(1+s\tau_c)}}\;,
\end{align}
and
\begin{align}
    F(q_n, s) ={} \frac{\tau s}{Lq_n} \int_{0}^{\infty}  dte^{-st}\left[ (-1)^n f_2(t) - f_1(t) \right]\;,
       \label{eq:Fqs}
\end{align}
with $A=\alpha\beta$ quantifying activity, as defined in the main text.

We now consider the specific form of the perturbation described in the main text, with $f_1(t)=-f_2(t)=\Delta \sin{(\omega_0t)}$, which represents symmetric oscillatory boundary displacements at frequency $\omega_0$. In this case by symmetry only Fourier component corresponding to even values of $n=2m$ are nonzero 
\begin{gather}
    \label{eq:vnsosc}
    \hat{v}_{2m}(s) = -\frac{4\Delta}{2m\pi}  \frac{b^2(s)}{(b^2(s) + q_{2m}^2)} \frac{\omega_0}{(s^2+\omega_0^2)}~.
\end{gather}

Inserting the explicit expression for $b(s)$, the inverse Laplace transform  is given by
\begin{widetext}
    \begin{align}
    \label{eq:vnpole}
    v_{2m}(t) ={}&- \frac{2\Delta\omega_0}{m\pi}
    \int_{s=\delta-i\infty}^{\delta+i\infty} \frac{ds}{2\pi i} \frac{s(1+s\tau_a)(1+s\tau_c)e^{st}}{(s^2+\omega_0^2) ( A q_{2m}^2+(s+q_{2m}^2)(1+s\tau_a)(1+s\tau_c))} ~.
    \end{align}
\end{widetext}

There are multiple poles in the denominator of Eq.~\eqref{eq:vnpole}: $s=\pm i\omega_0$, corresponding to the frequency of the external perturbation, and the poles of the susceptibility corresponding to $b^2(s)=-q_{2m}^2$.
The latter are simply given by the dispersion relations of the natural modes of the system evaluated in Section 2. One of these modes, denoted by $s_d(q_n)$ is always real and negative, hence decays over times scales of the order of $\tau$. The other two modes, with dispersion relation $s_{1,2}(q_n)$  can become complex and unstable for $A>A_{c}$. Thus $v_{2m}(t)$ can be written as
\begin{widetext}
    \begin{align}
    \label{eq:vnt}
    v_{2m}(t) ={}& -\frac{2\Delta\omega_0}{m\pi} \int_{\delta-i\infty}^{\delta+i\infty} \frac{ds}{2\pi i} \frac{s(1/\tau_a+s)(1/\tau_c+s)e^{st}}{(s-i\omega_0)(s+i\omega_0)[s-s_d(q_{2m})][s-s_1(q_{2m})][s-s_2(q_{2m})]}\;.
\end{align}
\end{widetext}

\begin{figure}[ht]    
    \includegraphics[width=0.45\textwidth]{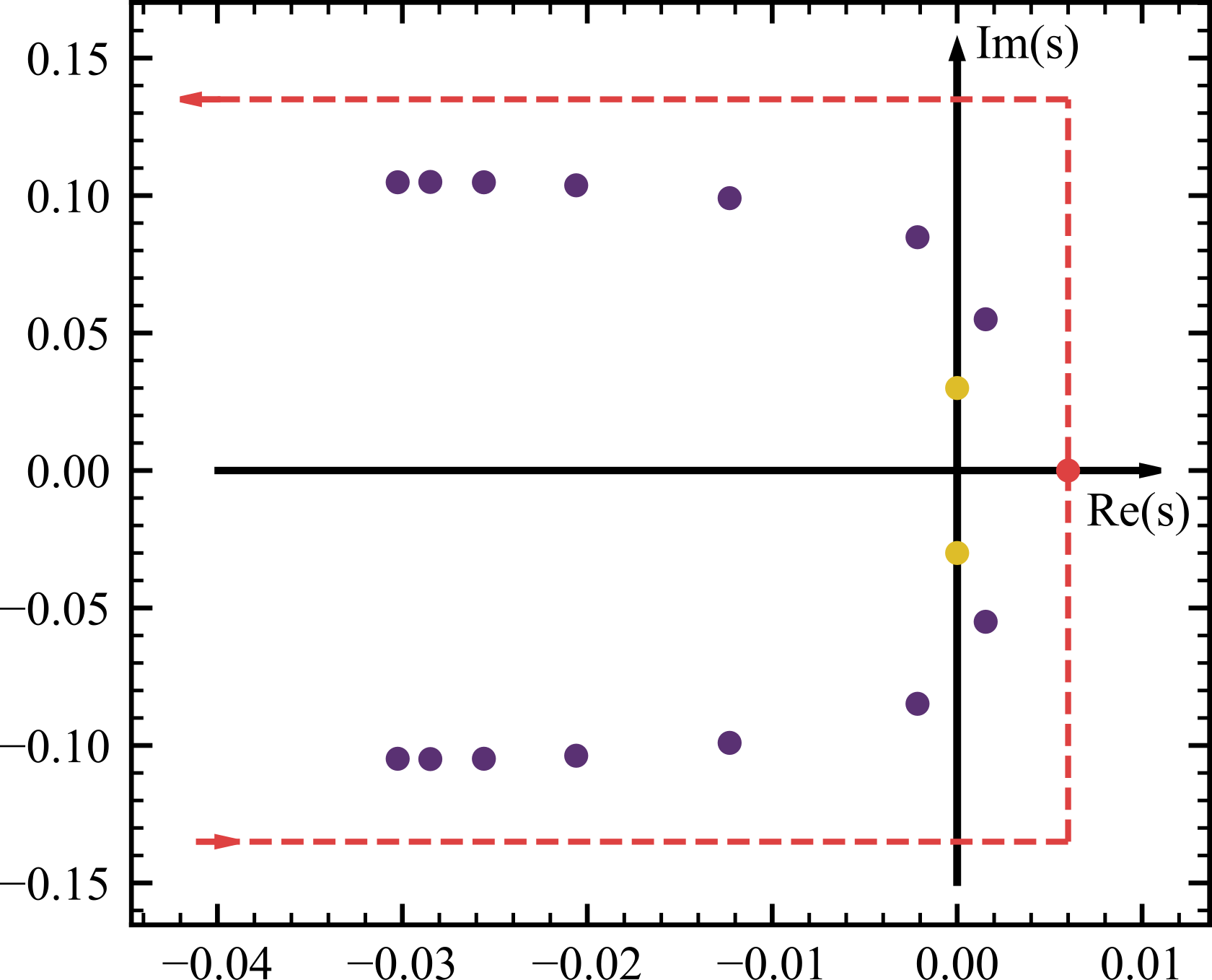}
    \caption{Plot of poles of Eq. \eqref{eq:vnpole} in complex plane for $A>A_c$. The purple dots are the first seven solutions to $b^2(s)=-q_{2m}^2$} for $m=1$ to $7$. The dots above x-axis are $s_1(q_{2m})$, the lower dots are $s_2(q_{2m})$, and $s_d(q_{2m})$ all lie on the left x-axis out of the field of view. The pair indexed by $m=1$ is to the right of the y axis and hence not a transient. The yellow dots are marginally stable perturbation modes $s=\pm i\omega_0$. The red dot is $\delta$, chosen such that it remains to the right of all poles and the contour is closed in a large half-circle to the left, depicted by the dashed red lines. Parameter values: $\tau_a/\tau=20, \tau_c/\tau=50, A_c=9.3, A=11.0$.
    \label{fig:pole_plot}
\end{figure}

We cannot obtain an explicit solution for $v_n(t)$ because the location of the modes in the imaginary plane depends on the value of $A$. We estimate the behavior for $A>A_c$ by including only the contribution from unstable modes, and evaluating those at the wavenumber corresponding to the fastest growth rate. For $A$ approaching  $A_c$ from above, the fastest growing wavenumber is $q_c = (\tau_a \tau_c)^{-1/4} $. Setting $q_c\sim \pi m/L$, we find that the mode corresponding to the fastest growing mode is $ m_c = \lfloor q_c L/\pi \rfloor$ ($\lfloor x\rfloor$ represents floor of $x$). An example of the mode structure is shown in Fig.~\ref{fig:pole_plot} that displays the first seven zeros of $b^2(s)+q_{2m}^2=0$ for $m=1,2,\cdots,8$. The purple dots denote $s_{1,2}(q_{2m})$, while $s_d(q_{2m})$ are all on the negative $x$-axis and out of the field of view. For the parameter values used in the figure, $m_c=1$ and only the pair indexed by $m=1$ is unstable. 
The integral in Eq.~\eqref{eq:vnt} is then calculated by closing the contour  in a large half-circle to negative $x$ (red path in Fig.~\ref{fig:pole_plot}) and including only the contribution from the most unstable modes and the zeros corresponding to the external frequency $\omega_0$.
All other modes are transient and do not affect long term behaviour. We note that for $A>A_c$ there is an entire band of growing modes corresponding to $\lfloor[q_-]<q_{2m}<\lfloor[q_+]$. 
Restricting ourselves to $A\sim A_c$ and setting 
$s_1(q_{2m_c})\equiv s_c$, $s_2(q_{2m_c})\equiv \bar s_c$  and $s_d\equiv s_d(q_{2m_c})$, where the bar denotes complex conjugate, 
we find
\begin{widetext}
    \begin{align}
        \begin{split}
        v(x,t) ={}& -\Delta~\sin{(\omega_0 t)}\left(1-\frac{x}{L}\right) + \Delta \text{Re}\left\{ie^{-i\omega_0 t} \left(\frac{\sinh{(b(-i\omega_0)(2L-x))}-\sinh{(b(-i\omega_0)x)}}{\sinh{(b(-i\omega_0)2L)}}\right)\right\}\\&
        -\frac{4\Delta\sin{(m_c \pi x/L)}}{m_c \pi} \text{Re} \left\{ \frac{s_c(1/\tau_a+s_c)(1/\tau_c+s_c)~e^{s_ct}}{(s_c^2+\omega_0^2)(s_c-s_d)(s_c-\bar s_c)} \right\} \\&
        + (\text{decaying modes})\;.
     \end{split}
    \end{align}
Finally, we can now write the expression for the displacement, $u(x,t)$ as
    \begin{align}
    \label{eq:uxt}
    u(x,t) ={}&  \Delta 
    \sin{\omega_0 t} ~\text{Re}[\mathcal{B}(x)]
    + \Delta 
    \cos{\omega_0 t} ~\text{Im}[\mathcal{B}(x)]
    -4\Delta\frac{\sin{(m_c \pi x/L)}}{m_c \pi} \text{Re} \left[ \frac{~e^{s_ct}}{(s_c^2+\omega_0^2) \mathcal{T}(s_c)} \right] + \text{decaying  modes}\;,
    \end{align}
    where
   \begin{align}
    \mathcal{B}(x) ={}& \frac{\sinh{(b(i\omega_0)(2L-x))}-\sinh{(b(i\omega_0)x)}}{\sinh{(2b(i\omega_0)L)}}\;,
    \\
     \mathcal{T}(s_c) ={}& \frac{1}{s_c} \left[1+\left(s_c+\frac{m_c^2 \pi^2}{L^2}\right)\left(\frac{1}{s_c+\tau_a^{-1}}+\frac{1}{s_c+\tau_c^{-1}}\right)\right]\;.
\end{align}
   \end{widetext}
Here $\mathcal{B}(x)$ encodes all spatial dependence and $\mathcal{T}(s_c)$ is an effective complex time scale.
In the main text $\mathcal{B}(x)$ is written in terms of a magnitude and a phase as 
\begin{align}
     \mathcal{B}(x)=|\mathcal{B}(x)|e^{i\phi_1}\;.
\end{align}
The explicit expression for $|\mathcal{B}(x)|$ and $\phi_1$ are not particularly illuminating and are not given here.

\section{Full Expressions for Inverse Perturbation length}
Below we write an expression for the squared inverse penetration length, $b^2(i\omega_0)$, with explicit real and imaginary parts. The expression involves growthrates evaluated at $q_\pm$, $s(q_\pm)=i\omega_\pm$, which are purely imaginary by definition. We find the value of $\omega_\pm$ in the same calculation as suggested for $q_\pm$. Here we draw attention to two facts, the real part of the squared inverse perturbation length $\text{Re}(b^2(i\omega_0))$ is always negative for all perturbation frequencies. Secondly, the imaginary part $\text{Im}(b^2(i\omega_0))$ vanishes at $\omega_\pm$. Thus the penetration length, defined as $l_p=1/\text{Re}(b(i\omega_0))$ diverges at $\omega_\pm$. It can also be seen that at resonance (for $\omega_0=\omega_c$, which lies between $\omega_\pm$) the imaginary part of squared inverse penetration length is small, leading to a large penetration length.

\begin{widetext}
    \begin{align}
    b^2(i\omega_0) &={} \frac{-\omega_0^2\tau(\tau_a+\tau_c)A + i \omega_0\tau \tau_a^2\tau_c^2(\omega_0^2-\omega_+^2)(\omega_0^2-\omega_-^2)}{\omega_0^2(\tau_a+\tau_c)^2 + (1+A-\omega_0^2\tau_a\tau_c)^2}\\
    \ell_p = 1/\text{Re}(b(i\omega_0)) &={} \sqrt{\frac{2}{\text{Re}(b^2) + |b^2| }}
    \text{, for } \text{Im}(b^2) << \text{Re}(b^2)~, \ell_p \simeq \frac{\sqrt{|\text{Re}(b^2)|}}{|\text{Im}(b^2)|}\\
    s(q_\pm) = i\omega_\pm  &={} i\frac{A\tau_a\tau_c-(\tau_a^2+\tau_c^2)\pm \sqrt{(\tau_a^2+\tau_c^2-A\tau_a\tau_c)^2-4\tau_a^2\tau_c^2(1+A)}}{2\tau_a^2\tau_c^2}\\
    \end{align}
\end{widetext}

\section{Effect of Phase of Perturbation}
In this section we include the effect of phase of applied perturbation on the final solution.
We find that the phase at which the perturbation was applied directly enters the natural amplitude term. With the perturbation $u(0,t)=-u(2L,t)=\Delta \sin(\omega_0t+\phi_0)$, the solution to displacement field is given by Eq. \eqref{eq:uxtphi}.
\begin{widetext}
    \begin{align}
     \label{eq:uxtphi}
    u(x,t) &={}  \Delta |\mathcal{B}(x)|
    \sin{(\omega_0 t+\phi_0+\phi_1)} -\frac{4\Delta\sin{(m_c\pi x/L)}}{m_c\pi} \text{Re} \left\{ \frac{(\omega_0\cos(\phi_0)+s_c\sin(\phi_0))~e^{s_ct}}{(s_c^2-(i\omega_0)^2) \mathcal{T}(s_c)} \right\}
\end{align}
\end{widetext}

\section{Periodic Boundary Conditions}
Changing the boundary conditions to periodic, we try to compare our results with experiments on ring shaped geometries. When starting from a homogeneous state, we first see standing waves develop, which then become travelling waves. The direction of travel is chosen spontaneously and the speed of the travelling wave is determined by the highest unstable wave-vector and its oscillation frequency. We see that the condition for a wave becomes $n\lambda=N\lbar$. More importantly, in the current model, we cannot see global rotations, as is seen experimentally in \cite{LoVecchio2024SpontaneousBoundaries,Jain2020}. To study such behaviour we would need to include cell-polarity in our model.

\section{Response to a square wave perturbation}

A square wave of period $T_0=2\pi/\omega_0$ and amplitude $\Delta$ can be written as the sum of a Fourier series: $\frac{4\Delta}{\pi} \sum_{i=1}^\infty \frac{1}{(2n-1)}\sin\left((2n-1)\omega_0 t\right)$. The first term, which is of the highest amplitude is the sinusoidal perturbation of frequency $\omega_0$. Thus, substituting a square wave in place of a sine wave will qualitatively give the same results. There will be a few differences, there might be additional peaks in resonance where multiples of the external frequency might equal the natural frequency. The second largest mode with frequency $3\omega_0$ is most likely to appear. We see this peak at $\omega_0=\omega_c/3$ in some simulations.
\begin{figure}[ht]
    \centering  
    \includegraphics[width=0.45\textwidth]{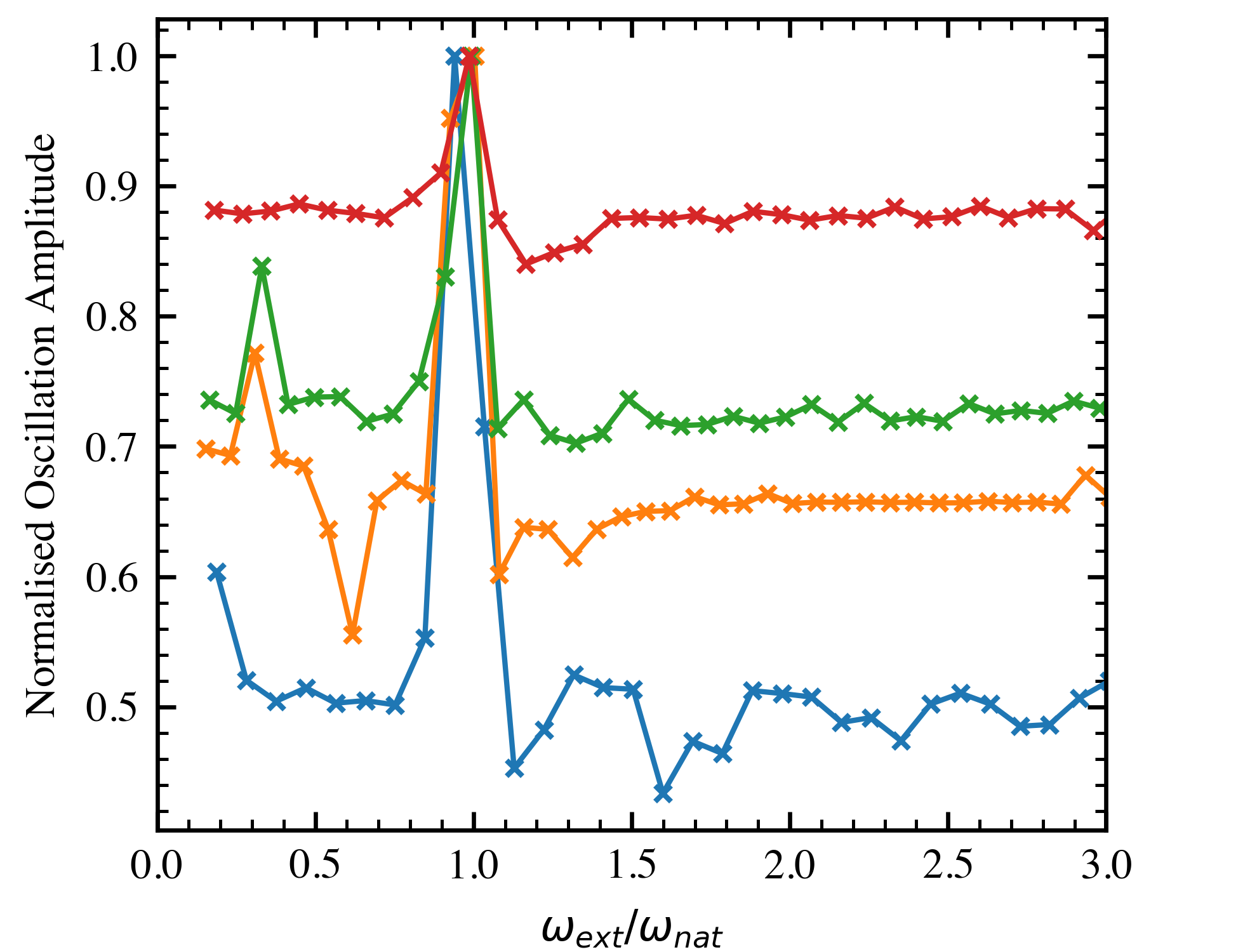}
    \caption{Resonance for 1-D cell trains of different lengths (Red: $L=1500 \mu m$, Green: $L=1000\mu m$, Yellow: $L=600 \mu m$, Blue: $L=400\mu m$). There's a peak when external frequency equals the natural frequency $\omega_c$ and since there are higher harmonics in a square wave we also get the peaks at $\omega_c/3$.}
    \label{fig:resonance_square}
\end{figure}
\begin{figure}[ht]
    \centering  
    \includegraphics[width=0.45\textwidth]{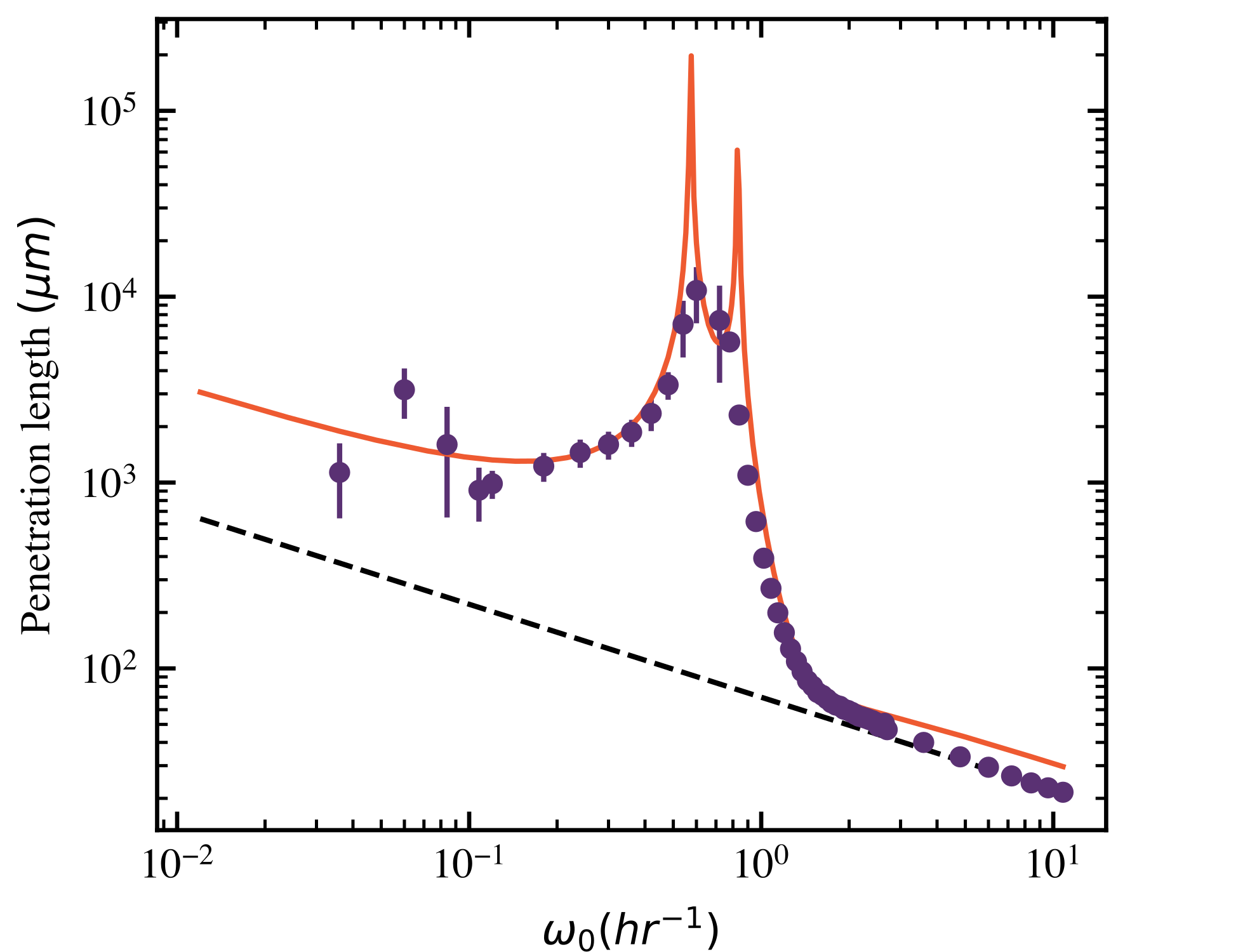}
    \caption{Penetration length for a square wave perturbation of different frequencies.}
    \label{fig:plen_square}
\end{figure}

\section{Step-stretch experimental methods}

\subsection{Cell culture}
Madin-Darby Canine Kidney (MDCK) II type G cells expressing a GFP E-cadherin reporter were cultured in low glucose DMEM (Thermofisher 11885084) supplemented with 10\% FBS and 1\% Penicillin Streptomycin at $37^\circ$C with 5\% CO2. Approximately 150,000 cells were seeded within the 35 millimeter glass bottom dish containing the protein patterns 1-2 days before line pulling experiments. Prior to experiments, cells were washed thoroughly with PBS to remove any cell debris or unadhered cells. Cells were imaged in phenol red free homemade DMEM (see Supplementary) supplemented with 50 mM HEPES, 10\% FBS, and 1\% Penicillin Streptomycin.

\subsection{Protein micropatterning}

35 mm glass bottom dishes (Cellvis, D35-20-1.5-N) were plasma treated for 5 minutes using atmospheric plasma at 18W (Harrick, PDC-32G). A custom cut PDMS ring (8 mm inner diameter, 12 mm outer diameter) was immediately placed in the center of the glass bottom and a solution of 100 µg/mL of poly(l-lysine)-graft-poly(ethylene glycol) (PLL(20)-g[3.5]- PEG(2)) (SuSoS AG, Dübendorf, Switzerland) diluted in PBS was added to the ring and left to incubate for 1h at room temperature. The PLL-g-PEG was then rinsed 10x with PBS with excess PBS removed. 20 µL of photoinitiator (PLPP, Alvéole) was added to the ring and the sample was then placed on the stage of a Leica DMi8 epifluorescence microscope equipped with a Fluotar 20x/0.40 N.A. objective and the Alvéole Primo photopatterning system (Alvéole, Paris, France). To create collagen lines wide enough for a single cell and long enough to capture long-range force propagation within our camera field of view, we designed patterns that were 20 $\mu$m wide by 390 $\mu$m long. Digital masks for collagen lines were made using the open-source software program Inkscape56 with a conversion determined by Primo calibration (0.277 $\mu$m/px). The Leonardo plugin (Alvéole Laboratory) on $\mu$manager software57 was used to create an array of digital masks, yielding 30 collagen lines per single glass bottom dish. Digital masks were projected through the glass at a dosage of 800 mJ/mm2 from a 375 nm, 7.10 mW laser.  Following light exposure, the photoinitiator was rinsed off with PBS and approximately 100 µL of 50 µg/mL rat tail collagen 1 (Corning, 354236) dissolved in 0.1 acetic acid was added to the ring. The sample was left to incubate for 1 hour at room temperature before finally being rinsed and stored at $4^\circ$C with PBS.

\subsection{Preparation of custom probes}
20 $\mu$m inner diameter prepulled glass micropipette tips (World Precision Instruments, TIP20TW1) were angled to approximately 120 degrees using an open flame for better fixture within the micromanipulator. A glass bottom polystyrene dish was then coated with optical glue (Norland Products, NOA81) and 30 $\mu$m diameter borosilicate glass beads (Cospheric, BSGMS-2.2 27-32 $\mu$m). Using a micromanipulator (Sensapex, uMp-3), the probe was delicately lowered in the optical glue and pressed gently against a glass bead. The probes were then pre-cured using a handheld UV laser (Lightfe, UV301 365 nm), and fully cured using a 365 nm UV lamp (Spectroline, EN-180). After the bead was glued to the glass micropipette tip, probes were plasma treated using atmospheric plasma at 18W (Harrick, PDC-32G), and immediately incubated for 1-3 hours at room temperature with 50 µg/mL rat tail collagen 1 (Corning, 354236) dissolved in 0.1\% acetic acid. The probe was then rinsed with phosphate buffered saline (PBS) and mounted to a micromanipulator (Sensapex, uMp-3) using a probe holder (Digitimer, H-12)

\subsection{Micromanipulation and Microscopy}
The sample containing cell trains was placed on the stage of a Zeiss Axio Observer 7 widefield fluorescence microscope placed on a vibration isolation table (Newport). To ensure cell viability, the cell media was supplemented with HEPES to buffer acidity and the microscope chamber was heated to $37^\circ$ C. The probe was lowered into the media containing dish, 1-2 $\mu$m above the substrate surface.  Mineral oil (Fisher Scientific, 8042-47-5) was added to the top of the dish to seal media and prevent evaporation. After the system had stabilized from thermal drift (1-3 hours), the probe was brought flush against the outermost cell of the cell train and left to form a strong stable attachment over 4-12 hours. The micromanipulator then applied a predetermined displacement of 30 $\mu$m at a rate of 1 $\mu$m/sec and simultaneously imaged every 2 seconds. Immediately following mechanical stretch, the cell line was imaged every 5 minutes for 6 hours with a 20x objective using brightfield imaging.

\subsection{Particle Image Velocimetry}
Raw brightfield images from before and after boundary strain were cropped to the confines of the cell train, excluding movement of the bead (approximately 400 $\mu$m x 25 $\mu$m). Image stacks were then exported as tiff files and analyzed using PIVLab (MATLAB, The MathWorks) with a sequence style of 1-2, 2-3, 3-4, etc.  For analysis, Fast Fourier Transform (FFT) window deformation was selected with a Gauss 2x3-point sub-pixel estimator. The cropped region was further refined with an ROI to yield 4-7 rows of ROI windows (i.e., 4-7 vectors in the vertical direction). For post-processing, parameters under vector validation included 4 as the standard deviation filter and 5 as the local median filter threshold. Vectors were then thresholded to exclude x velocities greater than 20 px/frame and less than -20 px/frame. Velocities in the y direction greater than 10 px/frame and less than -10 px/frame were also excluded.

\bibliography{references}

\begin{thebibliography}{45}%
\makeatletter
\providecommand \@ifxundefined [1]{%
 \@ifx{#1\undefined}
}%
\providecommand \@ifnum [1]{%
 \ifnum #1\expandafter \@firstoftwo
 \else \expandafter \@secondoftwo
 \fi
}%
\providecommand \@ifx [1]{%
 \ifx #1\expandafter \@firstoftwo
 \else \expandafter \@secondoftwo
 \fi
}%
\providecommand \natexlab [1]{#1}%
\providecommand \enquote  [1]{``#1''}%
\providecommand \bibnamefont  [1]{#1}%
\providecommand \bibfnamefont [1]{#1}%
\providecommand \citenamefont [1]{#1}%
\providecommand \href@noop [0]{\@secondoftwo}%
\providecommand \href [0]{\begingroup \@sanitize@url \@href}%
\providecommand \@href[1]{\@@startlink{#1}\@@href}%
\providecommand \@@href[1]{\endgroup#1\@@endlink}%
\providecommand \@sanitize@url [0]{\catcode `\\12\catcode `\$12\catcode `\&12\catcode `\#12\catcode `\^12\catcode `\_12\catcode `\%12\relax}%
\providecommand \@@startlink[1]{}%
\providecommand \@@endlink[0]{}%
\providecommand \url  [0]{\begingroup\@sanitize@url \@url }%
\providecommand \@url [1]{\endgroup\@href {#1}{\urlprefix }}%
\providecommand \urlprefix  [0]{URL }%
\providecommand \Eprint [0]{\href }%
\providecommand \doibase [0]{http://dx.doi.org/}%
\providecommand \selectlanguage [0]{\@gobble}%
\providecommand \bibinfo  [0]{\@secondoftwo}%
\providecommand \bibfield  [0]{\@secondoftwo}%
\providecommand \translation [1]{[#1]}%
\providecommand \BibitemOpen [0]{}%
\providecommand \bibitemStop [0]{}%
\providecommand \bibitemNoStop [0]{.\EOS\space}%
\providecommand \EOS [0]{\spacefactor3000\relax}%
\providecommand \BibitemShut  [1]{\csname bibitem#1\endcsname}%
\let\auto@bib@innerbib\@empty
\bibitem [{\citenamefont {Petrolli}\ \emph {et~al.}(2021)\citenamefont {Petrolli}, \citenamefont {Boudou}, \citenamefont {Balland},\ and\ \citenamefont {Cappello}}]{Petrolli2021OscillationsMigration}%
  \BibitemOpen
  \bibfield  {author} {\bibinfo {author} {\bibfnamefont {V.}~\bibnamefont {Petrolli}}, \bibinfo {author} {\bibfnamefont {T.}~\bibnamefont {Boudou}}, \bibinfo {author} {\bibfnamefont {M.}~\bibnamefont {Balland}}, \ and\ \bibinfo {author} {\bibfnamefont {G.}~\bibnamefont {Cappello}},\ }\href {\doibase 10.1016/B978-0-12-820310-1.00004-5} {\bibfield  {journal} {\bibinfo  {journal} {Viscoelasticity and Collective Cell Migration: An Interdisciplinary Perspective Across Levels of Organization}\ ,\ \bibinfo {pages} {157}} (\bibinfo {year} {2021})}\BibitemShut {NoStop}%
\bibitem [{\citenamefont {De~Leon}\ \emph {et~al.}(2023)\citenamefont {De~Leon}, \citenamefont {Wen}, \citenamefont {Paylaga}, \citenamefont {Wang}, \citenamefont {Roan}, \citenamefont {Wang}, \citenamefont {Hsiao}, \citenamefont {Lin},\ and\ \citenamefont {Chen}}]{DeLeon2023MechanicalTailfin}%
  \BibitemOpen
  \bibfield  {author} {\bibinfo {author} {\bibfnamefont {M.~P.}\ \bibnamefont {De~Leon}}, \bibinfo {author} {\bibfnamefont {F.~L.}\ \bibnamefont {Wen}}, \bibinfo {author} {\bibfnamefont {G.~J.}\ \bibnamefont {Paylaga}}, \bibinfo {author} {\bibfnamefont {Y.~T.}\ \bibnamefont {Wang}}, \bibinfo {author} {\bibfnamefont {H.~Y.}\ \bibnamefont {Roan}}, \bibinfo {author} {\bibfnamefont {C.~H.}\ \bibnamefont {Wang}}, \bibinfo {author} {\bibfnamefont {C.~D.}\ \bibnamefont {Hsiao}}, \bibinfo {author} {\bibfnamefont {K.~H.}\ \bibnamefont {Lin}}, \ and\ \bibinfo {author} {\bibfnamefont {C.~H.}\ \bibnamefont {Chen}},\ }\href {\doibase 10.1038/s41567-023-02103-6} {\bibfield  {journal} {\bibinfo  {journal} {Nature Physics 2023 19:9}\ }\textbf {\bibinfo {volume} {19}},\ \bibinfo {pages} {1362} (\bibinfo {year} {2023})}\BibitemShut {NoStop}%
\bibitem [{\citenamefont {Armon}\ \emph {et~al.}(2021)\citenamefont {Armon}, \citenamefont {Bull}, \citenamefont {Moriel}, \citenamefont {Aharoni},\ and\ \citenamefont {Prakash}}]{Armon2021}%
  \BibitemOpen
  \bibfield  {author} {\bibinfo {author} {\bibfnamefont {S.}~\bibnamefont {Armon}}, \bibinfo {author} {\bibfnamefont {M.~S.}\ \bibnamefont {Bull}}, \bibinfo {author} {\bibfnamefont {A.}~\bibnamefont {Moriel}}, \bibinfo {author} {\bibfnamefont {H.}~\bibnamefont {Aharoni}}, \ and\ \bibinfo {author} {\bibfnamefont {M.}~\bibnamefont {Prakash}},\ }\href {\doibase 10.1038/s42005-021-00712-2} {\bibfield  {journal} {\bibinfo  {journal} {Communications Physics 2021 4:1}\ }\textbf {\bibinfo {volume} {4}},\ \bibinfo {pages} {1} (\bibinfo {year} {2021})}\BibitemShut {NoStop}%
\bibitem [{\citenamefont {Gong}\ \emph {et~al.}(2023)\citenamefont {Gong}, \citenamefont {van~den Dries}, \citenamefont {Migueles-Ram{\'{i}}rez}, \citenamefont {Wiseman}, \citenamefont {Cambi},\ and\ \citenamefont {Shenoy}}]{Gong2023Chemo-mechanicalPodosomes}%
  \BibitemOpen
  \bibfield  {author} {\bibinfo {author} {\bibfnamefont {Z.}~\bibnamefont {Gong}}, \bibinfo {author} {\bibfnamefont {K.}~\bibnamefont {van~den Dries}}, \bibinfo {author} {\bibfnamefont {R.~A.}\ \bibnamefont {Migueles-Ram{\'{i}}rez}}, \bibinfo {author} {\bibfnamefont {P.~W.}\ \bibnamefont {Wiseman}}, \bibinfo {author} {\bibfnamefont {A.}~\bibnamefont {Cambi}}, \ and\ \bibinfo {author} {\bibfnamefont {V.~B.}\ \bibnamefont {Shenoy}},\ }\href {\doibase 10.1038/s41467-023-38598-z} {\bibfield  {journal} {\bibinfo  {journal} {Nature Communications 2023 14:1}\ }\textbf {\bibinfo {volume} {14}},\ \bibinfo {pages} {1} (\bibinfo {year} {2023})}\BibitemShut {NoStop}%
\bibitem [{\citenamefont {Staddon}\ \emph {et~al.}(2019)\citenamefont {Staddon}, \citenamefont {Cavanaugh}, \citenamefont {Munro}, \citenamefont {Gardel},\ and\ \citenamefont {Banerjee}}]{Staddon2019MechanosensitiveMorphogenesis}%
  \BibitemOpen
  \bibfield  {author} {\bibinfo {author} {\bibfnamefont {M.~F.}\ \bibnamefont {Staddon}}, \bibinfo {author} {\bibfnamefont {K.~E.}\ \bibnamefont {Cavanaugh}}, \bibinfo {author} {\bibfnamefont {E.~M.}\ \bibnamefont {Munro}}, \bibinfo {author} {\bibfnamefont {M.~L.}\ \bibnamefont {Gardel}}, \ and\ \bibinfo {author} {\bibfnamefont {S.}~\bibnamefont {Banerjee}},\ }\href {\doibase 10.1016/j.bpj.2019.09.027} {\bibfield  {journal} {\bibinfo  {journal} {Biophysical Journal}\ }\textbf {\bibinfo {volume} {117}},\ \bibinfo {pages} {1739} (\bibinfo {year} {2019})}\BibitemShut {NoStop}%
\bibitem [{\citenamefont {Fang}\ \emph {et~al.}(2022)\citenamefont {Fang}, \citenamefont {Yao}, \citenamefont {Zhang},\ and\ \citenamefont {Lin}}]{Fang2022ActiveSurfaces}%
  \BibitemOpen
  \bibfield  {author} {\bibinfo {author} {\bibfnamefont {C.}~\bibnamefont {Fang}}, \bibinfo {author} {\bibfnamefont {J.}~\bibnamefont {Yao}}, \bibinfo {author} {\bibfnamefont {Y.}~\bibnamefont {Zhang}}, \ and\ \bibinfo {author} {\bibfnamefont {Y.}~\bibnamefont {Lin}},\ }\href {\doibase 10.1016/J.BPJ.2022.02.028} {\bibfield  {journal} {\bibinfo  {journal} {Biophysical Journal}\ }\textbf {\bibinfo {volume} {121}},\ \bibinfo {pages} {1266} (\bibinfo {year} {2022})}\BibitemShut {NoStop}%
\bibitem [{\citenamefont {Serra-Picamal}\ \emph {et~al.}(2012)\citenamefont {Serra-Picamal}, \citenamefont {Conte}, \citenamefont {Vincent}, \citenamefont {Anon}, \citenamefont {Tambe}, \citenamefont {Bazellieres}, \citenamefont {Butler}, \citenamefont {Fredberg},\ and\ \citenamefont {Trepat}}]{Serra-Picamal2012MechanicalExpansion}%
  \BibitemOpen
  \bibfield  {author} {\bibinfo {author} {\bibfnamefont {X.}~\bibnamefont {Serra-Picamal}}, \bibinfo {author} {\bibfnamefont {V.}~\bibnamefont {Conte}}, \bibinfo {author} {\bibfnamefont {R.}~\bibnamefont {Vincent}}, \bibinfo {author} {\bibfnamefont {E.}~\bibnamefont {Anon}}, \bibinfo {author} {\bibfnamefont {D.~T.}\ \bibnamefont {Tambe}}, \bibinfo {author} {\bibfnamefont {E.}~\bibnamefont {Bazellieres}}, \bibinfo {author} {\bibfnamefont {J.~P.}\ \bibnamefont {Butler}}, \bibinfo {author} {\bibfnamefont {J.~J.}\ \bibnamefont {Fredberg}}, \ and\ \bibinfo {author} {\bibfnamefont {X.}~\bibnamefont {Trepat}},\ }\href {\doibase 10.1038/NPHYS2355} {\  (\bibinfo {year} {2012}),\ 10.1038/NPHYS2355}\BibitemShut {NoStop}%
\bibitem [{\citenamefont {Tlili}\ \emph {et~al.}(2018)\citenamefont {Tlili}, \citenamefont {Gauquelin}, \citenamefont {Li}, \citenamefont {Cardoso}, \citenamefont {Ladoux}, \citenamefont {Ayari},\ and\ \citenamefont {Graner}}]{Tlili2018CollectiveVelocity}%
  \BibitemOpen
  \bibfield  {author} {\bibinfo {author} {\bibfnamefont {S.}~\bibnamefont {Tlili}}, \bibinfo {author} {\bibfnamefont {E.}~\bibnamefont {Gauquelin}}, \bibinfo {author} {\bibfnamefont {B.}~\bibnamefont {Li}}, \bibinfo {author} {\bibfnamefont {O.}~\bibnamefont {Cardoso}}, \bibinfo {author} {\bibfnamefont {B.}~\bibnamefont {Ladoux}}, \bibinfo {author} {\bibfnamefont {H.~D.}\ \bibnamefont {Ayari}}, \ and\ \bibinfo {author} {\bibfnamefont {F.}~\bibnamefont {Graner}},\ }\href {\doibase 10.1098/RSOS.172421} {\bibfield  {journal} {\bibinfo  {journal} {Royal Society Open Science}\ }\textbf {\bibinfo {volume} {5}} (\bibinfo {year} {2018}),\ 10.1098/RSOS.172421}\BibitemShut {NoStop}%
\bibitem [{\citenamefont {Boocock}\ \emph {et~al.}(2020)\citenamefont {Boocock}, \citenamefont {Hino}, \citenamefont {Ruzickova}, \citenamefont {Hirashima},\ and\ \citenamefont {Hannezo}}]{Boocock2020TheoryMonolayers}%
  \BibitemOpen
  \bibfield  {author} {\bibinfo {author} {\bibfnamefont {D.}~\bibnamefont {Boocock}}, \bibinfo {author} {\bibfnamefont {N.}~\bibnamefont {Hino}}, \bibinfo {author} {\bibfnamefont {N.}~\bibnamefont {Ruzickova}}, \bibinfo {author} {\bibfnamefont {T.}~\bibnamefont {Hirashima}}, \ and\ \bibinfo {author} {\bibfnamefont {E.}~\bibnamefont {Hannezo}},\ }\href {\doibase 10.1038/s41567-020-01037-7} {\bibfield  {journal} {\bibinfo  {journal} {Nature Physics 2020 17:2}\ }\textbf {\bibinfo {volume} {17}},\ \bibinfo {pages} {267} (\bibinfo {year} {2020})}\BibitemShut {NoStop}%
\bibitem [{\citenamefont {Deforet}\ \emph {et~al.}(2014)\citenamefont {Deforet}, \citenamefont {Hakim}, \citenamefont {Yevick}, \citenamefont {Duclos},\ and\ \citenamefont {Silberzan}}]{Deforet2014ARTICLEConfinement}%
  \BibitemOpen
  \bibfield  {author} {\bibinfo {author} {\bibfnamefont {M.}~\bibnamefont {Deforet}}, \bibinfo {author} {\bibfnamefont {V.}~\bibnamefont {Hakim}}, \bibinfo {author} {\bibfnamefont {H.~G.}\ \bibnamefont {Yevick}}, \bibinfo {author} {\bibfnamefont {G.}~\bibnamefont {Duclos}}, \ and\ \bibinfo {author} {\bibfnamefont {.~P.}\ \bibnamefont {Silberzan}},\ }\href {\doibase 10.1038/ncomms4747} {\  (\bibinfo {year} {2014}),\ 10.1038/ncomms4747}\BibitemShut {NoStop}%
\bibitem [{\citenamefont {Peyret}\ \emph {et~al.}(2019)\citenamefont {Peyret}, \citenamefont {Mueller}, \citenamefont {d'Alessandro}, \citenamefont {Begnaud}, \citenamefont {Marcq}, \citenamefont {M{\`{e}}ge}, \citenamefont {Yeomans}, \citenamefont {Doostmohammadi},\ and\ \citenamefont {Ladoux}}]{Peyret2019SustainedSheets}%
  \BibitemOpen
  \bibfield  {author} {\bibinfo {author} {\bibfnamefont {G.}~\bibnamefont {Peyret}}, \bibinfo {author} {\bibfnamefont {R.}~\bibnamefont {Mueller}}, \bibinfo {author} {\bibfnamefont {J.}~\bibnamefont {d'Alessandro}}, \bibinfo {author} {\bibfnamefont {S.}~\bibnamefont {Begnaud}}, \bibinfo {author} {\bibfnamefont {P.}~\bibnamefont {Marcq}}, \bibinfo {author} {\bibfnamefont {R.~M.}\ \bibnamefont {M{\`{e}}ge}}, \bibinfo {author} {\bibfnamefont {J.~M.}\ \bibnamefont {Yeomans}}, \bibinfo {author} {\bibfnamefont {A.}~\bibnamefont {Doostmohammadi}}, \ and\ \bibinfo {author} {\bibfnamefont {B.}~\bibnamefont {Ladoux}},\ }\href {\doibase 10.1016/J.BPJ.2019.06.013} {\bibfield  {journal} {\bibinfo  {journal} {Biophysical Journal}\ }\textbf {\bibinfo {volume} {117}},\ \bibinfo {pages} {464} (\bibinfo {year} {2019})}\BibitemShut {NoStop}%
\bibitem [{\citenamefont {Petrolli}\ \emph {et~al.}(2019)\citenamefont {Petrolli}, \citenamefont {Le~Goff}, \citenamefont {Tadrous}, \citenamefont {Martens}, \citenamefont {Allier}, \citenamefont {Mandula}, \citenamefont {Herv{\'{e}}}, \citenamefont {Henkes}, \citenamefont {Sknepnek}, \citenamefont {Boudou}, \citenamefont {Cappello},\ and\ \citenamefont {Balland}}]{Petrolli2019Confinement-InducedModes}%
  \BibitemOpen
  \bibfield  {author} {\bibinfo {author} {\bibfnamefont {V.}~\bibnamefont {Petrolli}}, \bibinfo {author} {\bibfnamefont {M.}~\bibnamefont {Le~Goff}}, \bibinfo {author} {\bibfnamefont {M.}~\bibnamefont {Tadrous}}, \bibinfo {author} {\bibfnamefont {K.}~\bibnamefont {Martens}}, \bibinfo {author} {\bibfnamefont {C.}~\bibnamefont {Allier}}, \bibinfo {author} {\bibfnamefont {O.}~\bibnamefont {Mandula}}, \bibinfo {author} {\bibfnamefont {L.}~\bibnamefont {Herv{\'{e}}}}, \bibinfo {author} {\bibfnamefont {S.}~\bibnamefont {Henkes}}, \bibinfo {author} {\bibfnamefont {R.}~\bibnamefont {Sknepnek}}, \bibinfo {author} {\bibfnamefont {T.}~\bibnamefont {Boudou}}, \bibinfo {author} {\bibfnamefont {G.}~\bibnamefont {Cappello}}, \ and\ \bibinfo {author} {\bibfnamefont {M.}~\bibnamefont {Balland}},\ }\href {\doibase 10.1103/PHYSREVLETT.122.168101/FIGURES/3/MEDIUM} {\bibfield  {journal} {\bibinfo  {journal} {Physical Review Letters}\ }\textbf {\bibinfo {volume} {122}},\ \bibinfo {pages} {168101} (\bibinfo {year}
  {2019})}\BibitemShut {NoStop}%
\bibitem [{\citenamefont {Cai}\ \emph {et~al.}(2014)\citenamefont {Cai}, \citenamefont {Chen}, \citenamefont {Prasad}, \citenamefont {He}, \citenamefont {Wang}, \citenamefont {Choesmel-Cadamuro}, \citenamefont {Sawyer}, \citenamefont {Danuser},\ and\ \citenamefont {Montell}}]{Cai2014MechanicalMigration}%
  \BibitemOpen
  \bibfield  {author} {\bibinfo {author} {\bibfnamefont {D.}~\bibnamefont {Cai}}, \bibinfo {author} {\bibfnamefont {S.~C.}\ \bibnamefont {Chen}}, \bibinfo {author} {\bibfnamefont {M.}~\bibnamefont {Prasad}}, \bibinfo {author} {\bibfnamefont {L.}~\bibnamefont {He}}, \bibinfo {author} {\bibfnamefont {X.}~\bibnamefont {Wang}}, \bibinfo {author} {\bibfnamefont {V.}~\bibnamefont {Choesmel-Cadamuro}}, \bibinfo {author} {\bibfnamefont {J.~K.}\ \bibnamefont {Sawyer}}, \bibinfo {author} {\bibfnamefont {G.}~\bibnamefont {Danuser}}, \ and\ \bibinfo {author} {\bibfnamefont {D.~J.}\ \bibnamefont {Montell}},\ }\href {\doibase 10.1016/j.cell.2014.03.045} {\bibfield  {journal} {\bibinfo  {journal} {Cell}\ }\textbf {\bibinfo {volume} {157}},\ \bibinfo {pages} {1146} (\bibinfo {year} {2014})}\BibitemShut {NoStop}%
\bibitem [{\citenamefont {Banerjee}\ \emph {et~al.}(2015)\citenamefont {Banerjee}, \citenamefont {Utuje},\ and\ \citenamefont {Marchetti}}]{Banerjee2015PropagatingExpansion}%
  \BibitemOpen
  \bibfield  {author} {\bibinfo {author} {\bibfnamefont {S.}~\bibnamefont {Banerjee}}, \bibinfo {author} {\bibfnamefont {K.~J.}\ \bibnamefont {Utuje}}, \ and\ \bibinfo {author} {\bibfnamefont {M.~C.}\ \bibnamefont {Marchetti}},\ }\href {\doibase 10.1103/PHYSREVLETT.114.228101/FIGURES/4/MEDIUM} {\bibfield  {journal} {\bibinfo  {journal} {Physical Review Letters}\ }\textbf {\bibinfo {volume} {114}},\ \bibinfo {pages} {228101} (\bibinfo {year} {2015})}\BibitemShut {NoStop}%
\bibitem [{\citenamefont {Notbohm}\ \emph {et~al.}(2016)\citenamefont {Notbohm}, \citenamefont {Banerjee}, \citenamefont {Utuje}, \citenamefont {Gweon}, \citenamefont {Jang}, \citenamefont {Park}, \citenamefont {Shin}, \citenamefont {Butler}, \citenamefont {Fredberg},\ and\ \citenamefont {Marchetti}}]{Notbohm2016CellularMotion}%
  \BibitemOpen
  \bibfield  {author} {\bibinfo {author} {\bibfnamefont {J.}~\bibnamefont {Notbohm}}, \bibinfo {author} {\bibfnamefont {S.}~\bibnamefont {Banerjee}}, \bibinfo {author} {\bibfnamefont {K.~J.}\ \bibnamefont {Utuje}}, \bibinfo {author} {\bibfnamefont {B.}~\bibnamefont {Gweon}}, \bibinfo {author} {\bibfnamefont {H.}~\bibnamefont {Jang}}, \bibinfo {author} {\bibfnamefont {Y.}~\bibnamefont {Park}}, \bibinfo {author} {\bibfnamefont {J.}~\bibnamefont {Shin}}, \bibinfo {author} {\bibfnamefont {J.~P.}\ \bibnamefont {Butler}}, \bibinfo {author} {\bibfnamefont {J.~J.}\ \bibnamefont {Fredberg}}, \ and\ \bibinfo {author} {\bibfnamefont {M.~C.}\ \bibnamefont {Marchetti}},\ }\href {\doibase 10.1016/J.BPJ.2016.05.019} {\bibfield  {journal} {\bibinfo  {journal} {Biophysical Journal}\ }\textbf {\bibinfo {volume} {110}},\ \bibinfo {pages} {2729} (\bibinfo {year} {2016})}\BibitemShut {NoStop}%
\bibitem [{\citenamefont {Banerjee}\ and\ \citenamefont {Marchetti}(2019)}]{Banerjee2019ContinuumMigration}%
  \BibitemOpen
  \bibfield  {author} {\bibinfo {author} {\bibfnamefont {S.}~\bibnamefont {Banerjee}}\ and\ \bibinfo {author} {\bibfnamefont {M.~C.}\ \bibnamefont {Marchetti}},\ }\href {\doibase 10.1007/978-3-030-17593-1{\_}4/FIGURES/10} {\bibfield  {journal} {\bibinfo  {journal} {Advances in Experimental Medicine and Biology}\ }\textbf {\bibinfo {volume} {1146}},\ \bibinfo {pages} {45} (\bibinfo {year} {2019})}\BibitemShut {NoStop}%
\bibitem [{\citenamefont {Alert}\ and\ \citenamefont {Trepat}(2020)}]{Alert2020PhysicalMigration}%
  \BibitemOpen
  \bibfield  {author} {\bibinfo {author} {\bibfnamefont {R.}~\bibnamefont {Alert}}\ and\ \bibinfo {author} {\bibfnamefont {X.}~\bibnamefont {Trepat}},\ }\href {\doibase 10.1146/ANNUREV-CONMATPHYS-031218-013516} {\bibfield  {journal} {\bibinfo  {journal} {https://doi.org/10.1146/annurev-conmatphys-031218-013516}\ }\textbf {\bibinfo {volume} {11}},\ \bibinfo {pages} {77} (\bibinfo {year} {2020})}\BibitemShut {NoStop}%
\bibitem [{\citenamefont {Vincent}\ \emph {et~al.}(2015)\citenamefont {Vincent}, \citenamefont {Bazelli{\`{e}}res}, \citenamefont {P{\'{e}}rez-Gonz{\'{a}}lez}, \citenamefont {Uroz}, \citenamefont {Serra-Picamal},\ and\ \citenamefont {Trepat}}]{Vincent2015ActiveMonolayer}%
  \BibitemOpen
  \bibfield  {author} {\bibinfo {author} {\bibfnamefont {R.}~\bibnamefont {Vincent}}, \bibinfo {author} {\bibfnamefont {E.}~\bibnamefont {Bazelli{\`{e}}res}}, \bibinfo {author} {\bibfnamefont {C.}~\bibnamefont {P{\'{e}}rez-Gonz{\'{a}}lez}}, \bibinfo {author} {\bibfnamefont {M.}~\bibnamefont {Uroz}}, \bibinfo {author} {\bibfnamefont {X.}~\bibnamefont {Serra-Picamal}}, \ and\ \bibinfo {author} {\bibfnamefont {X.}~\bibnamefont {Trepat}},\ }\href {\doibase 10.1103/PhysRevLett.115.248103} {\  (\bibinfo {year} {2015}),\ 10.1103/PhysRevLett.115.248103}\BibitemShut {NoStop}%
\bibitem [{\citenamefont {Safa}\ \emph {et~al.}(2024)\citenamefont {Safa}, \citenamefont {Huang}, \citenamefont {Kabla},\ and\ \citenamefont {Yang}}]{Safa2024ActiveMechanics}%
  \BibitemOpen
  \bibfield  {author} {\bibinfo {author} {\bibfnamefont {B.~T.}\ \bibnamefont {Safa}}, \bibinfo {author} {\bibfnamefont {C.}~\bibnamefont {Huang}}, \bibinfo {author} {\bibfnamefont {A.}~\bibnamefont {Kabla}}, \ and\ \bibinfo {author} {\bibfnamefont {R.}~\bibnamefont {Yang}},\ }\href {\doibase 10.1098/RSOS.231074} {\bibfield  {journal} {\bibinfo  {journal} {Royal Society Open Science}\ }\textbf {\bibinfo {volume} {11}} (\bibinfo {year} {2024}),\ 10.1098/RSOS.231074}\BibitemShut {NoStop}%
\bibitem [{\citenamefont {Bailles}\ \emph {et~al.}(2022)\citenamefont {Bailles}, \citenamefont {Gehrels},\ and\ \citenamefont {Lecuit}}]{Bailles2022MechanochemicalTissues}%
  \BibitemOpen
  \bibfield  {author} {\bibinfo {author} {\bibfnamefont {A.}~\bibnamefont {Bailles}}, \bibinfo {author} {\bibfnamefont {E.~W.}\ \bibnamefont {Gehrels}}, \ and\ \bibinfo {author} {\bibfnamefont {T.}~\bibnamefont {Lecuit}},\ }\href {\doibase 10.1146/annurev-cellbio-120420} {\  (\bibinfo {year} {2022}),\ 10.1146/annurev-cellbio-120420}\BibitemShut {NoStop}%
\bibitem [{\citenamefont {Khalilgharibi}\ \emph {et~al.}(2019)\citenamefont {Khalilgharibi}, \citenamefont {Fouchard}, \citenamefont {Asadipour}, \citenamefont {Barrientos}, \citenamefont {Duda}, \citenamefont {Bonfanti}, \citenamefont {Yonis}, \citenamefont {Harris}, \citenamefont {Mosaffa}, \citenamefont {Fujita}, \citenamefont {Kabla}, \citenamefont {Mao}, \citenamefont {Baum}, \citenamefont {Mu{\~{n}}oz}, \citenamefont {Miodownik},\ and\ \citenamefont {Charras}}]{Khalilgharibi2019StressCortex}%
  \BibitemOpen
  \bibfield  {author} {\bibinfo {author} {\bibfnamefont {N.}~\bibnamefont {Khalilgharibi}}, \bibinfo {author} {\bibfnamefont {J.}~\bibnamefont {Fouchard}}, \bibinfo {author} {\bibfnamefont {N.}~\bibnamefont {Asadipour}}, \bibinfo {author} {\bibfnamefont {R.}~\bibnamefont {Barrientos}}, \bibinfo {author} {\bibfnamefont {M.}~\bibnamefont {Duda}}, \bibinfo {author} {\bibfnamefont {A.}~\bibnamefont {Bonfanti}}, \bibinfo {author} {\bibfnamefont {A.}~\bibnamefont {Yonis}}, \bibinfo {author} {\bibfnamefont {A.}~\bibnamefont {Harris}}, \bibinfo {author} {\bibfnamefont {P.}~\bibnamefont {Mosaffa}}, \bibinfo {author} {\bibfnamefont {Y.}~\bibnamefont {Fujita}}, \bibinfo {author} {\bibfnamefont {A.}~\bibnamefont {Kabla}}, \bibinfo {author} {\bibfnamefont {Y.}~\bibnamefont {Mao}}, \bibinfo {author} {\bibfnamefont {B.}~\bibnamefont {Baum}}, \bibinfo {author} {\bibfnamefont {J.~J.}\ \bibnamefont {Mu{\~{n}}oz}}, \bibinfo {author} {\bibfnamefont {M.}~\bibnamefont {Miodownik}}, \ and\ \bibinfo {author} {\bibfnamefont
  {G.}~\bibnamefont {Charras}},\ }\href {\doibase 10.1038/s41567-019-0516-6} {\bibfield  {journal} {\bibinfo  {journal} {Nature Physics 2019 15:8}\ }\textbf {\bibinfo {volume} {15}},\ \bibinfo {pages} {839} (\bibinfo {year} {2019})}\BibitemShut {NoStop}%
\bibitem [{\citenamefont {Sadeghipour}\ \emph {et~al.}(2018)\citenamefont {Sadeghipour}, \citenamefont {Garcia}, \citenamefont {Nelson},\ and\ \citenamefont {Pruitt}}]{Sadeghipour2018}%
  \BibitemOpen
  \bibfield  {author} {\bibinfo {author} {\bibfnamefont {E.}~\bibnamefont {Sadeghipour}}, \bibinfo {author} {\bibfnamefont {M.~A.}\ \bibnamefont {Garcia}}, \bibinfo {author} {\bibfnamefont {W.~J.}\ \bibnamefont {Nelson}}, \ and\ \bibinfo {author} {\bibfnamefont {B.~L.}\ \bibnamefont {Pruitt}},\ }\href {\doibase 10.7554/ELIFE.39640} {\bibfield  {journal} {\bibinfo  {journal} {eLife}\ }\textbf {\bibinfo {volume} {7}} (\bibinfo {year} {2018}),\ 10.7554/ELIFE.39640}\BibitemShut {NoStop}%
\bibitem [{\citenamefont {Bodenschatz}\ \emph {et~al.}(2022)\citenamefont {Bodenschatz}, \citenamefont {Ajmail}, \citenamefont {Skamrahl}, \citenamefont {Vache}, \citenamefont {Gottwald}, \citenamefont {Nehls},\ and\ \citenamefont {Janshoff}}]{Bodenschatz2022EpithelialStress}%
  \BibitemOpen
  \bibfield  {author} {\bibinfo {author} {\bibfnamefont {J.~F.}\ \bibnamefont {Bodenschatz}}, \bibinfo {author} {\bibfnamefont {K.}~\bibnamefont {Ajmail}}, \bibinfo {author} {\bibfnamefont {M.}~\bibnamefont {Skamrahl}}, \bibinfo {author} {\bibfnamefont {M.}~\bibnamefont {Vache}}, \bibinfo {author} {\bibfnamefont {J.}~\bibnamefont {Gottwald}}, \bibinfo {author} {\bibfnamefont {S.}~\bibnamefont {Nehls}}, \ and\ \bibinfo {author} {\bibfnamefont {A.}~\bibnamefont {Janshoff}},\ }\href {\doibase 10.1038/s42003-022-03809-8} {\bibfield  {journal} {\bibinfo  {journal} {Communications Biology 2022 5:1}\ }\textbf {\bibinfo {volume} {5}},\ \bibinfo {pages} {1} (\bibinfo {year} {2022})}\BibitemShut {NoStop}%
\bibitem [{\citenamefont {P{\"{u}}llen}\ \emph {et~al.}(2021)\citenamefont {P{\"{u}}llen}, \citenamefont {Konrad}, \citenamefont {Merkel},\ and\ \citenamefont {Hoffmann}}]{Pullen2021SkinEpithelia}%
  \BibitemOpen
  \bibfield  {author} {\bibinfo {author} {\bibfnamefont {R.}~\bibnamefont {P{\"{u}}llen}}, \bibinfo {author} {\bibfnamefont {J.}~\bibnamefont {Konrad}}, \bibinfo {author} {\bibfnamefont {R.}~\bibnamefont {Merkel}}, \ and\ \bibinfo {author} {\bibfnamefont {B.}~\bibnamefont {Hoffmann}},\ }\href {\doibase 10.3390/CELLS10071834/S1} {\bibfield  {journal} {\bibinfo  {journal} {Cells}\ }\textbf {\bibinfo {volume} {10}} (\bibinfo {year} {2021}),\ 10.3390/CELLS10071834/S1}\BibitemShut {NoStop}%
\bibitem [{\citenamefont {Harris}\ \emph {et~al.}(2012)\citenamefont {Harris}, \citenamefont {Peter}, \citenamefont {Bellis}, \citenamefont {Baum}, \citenamefont {Kabla},\ and\ \citenamefont {Charras}}]{Harris2012CharacterizingMonolayers}%
  \BibitemOpen
  \bibfield  {author} {\bibinfo {author} {\bibfnamefont {A.~R.}\ \bibnamefont {Harris}}, \bibinfo {author} {\bibfnamefont {L.}~\bibnamefont {Peter}}, \bibinfo {author} {\bibfnamefont {J.}~\bibnamefont {Bellis}}, \bibinfo {author} {\bibfnamefont {B.}~\bibnamefont {Baum}}, \bibinfo {author} {\bibfnamefont {A.~J.}\ \bibnamefont {Kabla}}, \ and\ \bibinfo {author} {\bibfnamefont {G.~T.}\ \bibnamefont {Charras}},\ }\href {\doibase 10.1073/PNAS.1213301109/-/DCSUPPLEMENTAL} {\bibfield  {journal} {\bibinfo  {journal} {Proceedings of the National Academy of Sciences of the United States of America}\ }\textbf {\bibinfo {volume} {109}},\ \bibinfo {pages} {16449} (\bibinfo {year} {2012})}\BibitemShut {NoStop}%
\bibitem [{\citenamefont {Borghi}\ \emph {et~al.}(2012)\citenamefont {Borghi}, \citenamefont {Sorokina}, \citenamefont {Shcherbakova}, \citenamefont {Weis}, \citenamefont {Pruitt}, \citenamefont {Nelson},\ and\ \citenamefont {Dunn}}]{Borghi2012}%
  \BibitemOpen
  \bibfield  {author} {\bibinfo {author} {\bibfnamefont {N.}~\bibnamefont {Borghi}}, \bibinfo {author} {\bibfnamefont {M.}~\bibnamefont {Sorokina}}, \bibinfo {author} {\bibfnamefont {O.~G.}\ \bibnamefont {Shcherbakova}}, \bibinfo {author} {\bibfnamefont {W.~I.}\ \bibnamefont {Weis}}, \bibinfo {author} {\bibfnamefont {B.~L.}\ \bibnamefont {Pruitt}}, \bibinfo {author} {\bibfnamefont {W.~J.}\ \bibnamefont {Nelson}}, \ and\ \bibinfo {author} {\bibfnamefont {A.~R.}\ \bibnamefont {Dunn}},\ }\href {\doibase 10.1073/PNAS.1204390109/SUPPL{\_}FILE/PNAS.201204390SI.PDF} {\bibfield  {journal} {\bibinfo  {journal} {Proceedings of the National Academy of Sciences of the United States of America}\ }\textbf {\bibinfo {volume} {109}},\ \bibinfo {pages} {12568} (\bibinfo {year} {2012})}\BibitemShut {NoStop}%
\bibitem [{\citenamefont {Kinoshita}\ \emph {et~al.}(2020)\citenamefont {Kinoshita}, \citenamefont {Hashimoto}, \citenamefont {Yasue}, \citenamefont {Suzuki}, \citenamefont {Cristea},\ and\ \citenamefont {Ueno}}]{Kinoshita2020MechanicalEmbryogenesis}%
  \BibitemOpen
  \bibfield  {author} {\bibinfo {author} {\bibfnamefont {N.}~\bibnamefont {Kinoshita}}, \bibinfo {author} {\bibfnamefont {Y.}~\bibnamefont {Hashimoto}}, \bibinfo {author} {\bibfnamefont {N.}~\bibnamefont {Yasue}}, \bibinfo {author} {\bibfnamefont {M.}~\bibnamefont {Suzuki}}, \bibinfo {author} {\bibfnamefont {I.~M.}\ \bibnamefont {Cristea}}, \ and\ \bibinfo {author} {\bibfnamefont {N.}~\bibnamefont {Ueno}},\ }\href {\doibase 10.1016/j.celrep.2020.02.074} {\bibfield  {journal} {\bibinfo  {journal} {CellReports}\ }\textbf {\bibinfo {volume} {30}},\ \bibinfo {pages} {3875} (\bibinfo {year} {2020})}\BibitemShut {NoStop}%
\bibitem [{\citenamefont {Trepat}\ \emph {et~al.}(2007)\citenamefont {Trepat}, \citenamefont {Deng}, \citenamefont {An}, \citenamefont {Navajas}, \citenamefont {Tschumperlin}, \citenamefont {Gerthoffer}, \citenamefont {Butler},\ and\ \citenamefont {Fredberg}}]{Trepat2007UniversalCell}%
  \BibitemOpen
  \bibfield  {author} {\bibinfo {author} {\bibfnamefont {X.}~\bibnamefont {Trepat}}, \bibinfo {author} {\bibfnamefont {L.}~\bibnamefont {Deng}}, \bibinfo {author} {\bibfnamefont {S.~S.}\ \bibnamefont {An}}, \bibinfo {author} {\bibfnamefont {D.}~\bibnamefont {Navajas}}, \bibinfo {author} {\bibfnamefont {D.~J.}\ \bibnamefont {Tschumperlin}}, \bibinfo {author} {\bibfnamefont {W.~T.}\ \bibnamefont {Gerthoffer}}, \bibinfo {author} {\bibfnamefont {J.~P.}\ \bibnamefont {Butler}}, \ and\ \bibinfo {author} {\bibfnamefont {J.~J.}\ \bibnamefont {Fredberg}},\ }\href {\doibase 10.1038/nature05824} {\bibfield  {journal} {\bibinfo  {journal} {Nature 2007 447:7144}\ }\textbf {\bibinfo {volume} {447}},\ \bibinfo {pages} {592} (\bibinfo {year} {2007})}\BibitemShut {NoStop}%
\bibitem [{\citenamefont {Vercruysse}\ \emph {et~al.}(2024)\citenamefont {Vercruysse}, \citenamefont {Br{\"{u}}ckner}, \citenamefont {G{\'{o}}mez-Gonz{\'{a}}lez}, \citenamefont {Remson}, \citenamefont {Luciano}, \citenamefont {Kalukula}, \citenamefont {Rossetti}, \citenamefont {Trepat}, \citenamefont {Hannezo},\ and\ \citenamefont {Gabriele}}]{Vercruysse2024NatureClusters}%
  \BibitemOpen
  \bibfield  {author} {\bibinfo {author} {\bibfnamefont {E.}~\bibnamefont {Vercruysse}}, \bibinfo {author} {\bibfnamefont {D.~B.}\ \bibnamefont {Br{\"{u}}ckner}}, \bibinfo {author} {\bibfnamefont {M.}~\bibnamefont {G{\'{o}}mez-Gonz{\'{a}}lez}}, \bibinfo {author} {\bibfnamefont {A.}~\bibnamefont {Remson}}, \bibinfo {author} {\bibfnamefont {M.}~\bibnamefont {Luciano}}, \bibinfo {author} {\bibfnamefont {Y.}~\bibnamefont {Kalukula}}, \bibinfo {author} {\bibfnamefont {L.}~\bibnamefont {Rossetti}}, \bibinfo {author} {\bibfnamefont {X.}~\bibnamefont {Trepat}}, \bibinfo {author} {\bibfnamefont {E.}~\bibnamefont {Hannezo}}, \ and\ \bibinfo {author} {\bibfnamefont {S.}~\bibnamefont {Gabriele}},\ }\href {\doibase 10.1038/s41567-024-02532-x} {\bibfield  {journal} {\bibinfo  {journal} {Nature Physics |}\ }\textbf {\bibinfo {volume} {20}},\ \bibinfo {pages} {1492} (\bibinfo {year} {2024})}\BibitemShut {NoStop}%
\bibitem [{\citenamefont {Rossetti}\ \emph {et~al.}(2024)\citenamefont {Rossetti}, \citenamefont {Grosser}, \citenamefont {Abenza}, \citenamefont {Valon}, \citenamefont {Roca-Cusachs}, \citenamefont {Alert},\ and\ \citenamefont {Trepat}}]{Rossetti2024OptogeneticMigration}%
  \BibitemOpen
  \bibfield  {author} {\bibinfo {author} {\bibfnamefont {L.}~\bibnamefont {Rossetti}}, \bibinfo {author} {\bibfnamefont {S.}~\bibnamefont {Grosser}}, \bibinfo {author} {\bibfnamefont {J.~F.}\ \bibnamefont {Abenza}}, \bibinfo {author} {\bibfnamefont {L.}~\bibnamefont {Valon}}, \bibinfo {author} {\bibfnamefont {P.}~\bibnamefont {Roca-Cusachs}}, \bibinfo {author} {\bibfnamefont {R.}~\bibnamefont {Alert}}, \ and\ \bibinfo {author} {\bibfnamefont {X.}~\bibnamefont {Trepat}},\ }\href {\doibase 10.1038/s41567-024-02600-2} {\bibfield  {journal} {\bibinfo  {journal} {Nature Physics 2024 20:10}\ }\textbf {\bibinfo {volume} {20}},\ \bibinfo {pages} {1659} (\bibinfo {year} {2024})}\BibitemShut {NoStop}%
\bibitem [{\citenamefont {Chen}\ \emph {et~al.}(1998)\citenamefont {Chen}, \citenamefont {Mrksich}, \citenamefont {Huang}, \citenamefont {Whitesides},\ and\ \citenamefont {Ingber}}]{Chen1998MicropatternedFunction}%
  \BibitemOpen
  \bibfield  {author} {\bibinfo {author} {\bibfnamefont {C.~S.}\ \bibnamefont {Chen}}, \bibinfo {author} {\bibfnamefont {M.}~\bibnamefont {Mrksich}}, \bibinfo {author} {\bibfnamefont {S.}~\bibnamefont {Huang}}, \bibinfo {author} {\bibfnamefont {G.~M.}\ \bibnamefont {Whitesides}}, \ and\ \bibinfo {author} {\bibfnamefont {D.~E.}\ \bibnamefont {Ingber}},\ }\href {\doibase 10.1021/BP980031M} {\bibfield  {journal} {\bibinfo  {journal} {Biotechnology progress}\ }\textbf {\bibinfo {volume} {14}},\ \bibinfo {pages} {356} (\bibinfo {year} {1998})}\BibitemShut {NoStop}%
\bibitem [{\citenamefont {Tang}\ \emph {et~al.}(2012)\citenamefont {Tang}, \citenamefont {Yakut~Ali},\ and\ \citenamefont {Saif}}]{Tang2012AGels}%
  \BibitemOpen
  \bibfield  {author} {\bibinfo {author} {\bibfnamefont {X.}~\bibnamefont {Tang}}, \bibinfo {author} {\bibfnamefont {M.}~\bibnamefont {Yakut~Ali}}, \ and\ \bibinfo {author} {\bibfnamefont {M.~T.~A.}\ \bibnamefont {Saif}},\ }\href {\doibase 10.1039/C2SM25533B} {\bibfield  {journal} {\bibinfo  {journal} {Soft Matter}\ }\textbf {\bibinfo {volume} {8}},\ \bibinfo {pages} {7197} (\bibinfo {year} {2012})}\BibitemShut {NoStop}%
\bibitem [{\citenamefont {Moeller}\ \emph {et~al.}(2018)\citenamefont {Moeller}, \citenamefont {Denisin}, \citenamefont {Sim}, \citenamefont {Wilson}, \citenamefont {Ribeiro},\ and\ \citenamefont {Pruitt}}]{Moeller2018ControllingPatterning}%
  \BibitemOpen
  \bibfield  {author} {\bibinfo {author} {\bibfnamefont {J.}~\bibnamefont {Moeller}}, \bibinfo {author} {\bibfnamefont {A.~K.}\ \bibnamefont {Denisin}}, \bibinfo {author} {\bibfnamefont {J.~Y.}\ \bibnamefont {Sim}}, \bibinfo {author} {\bibfnamefont {R.~E.}\ \bibnamefont {Wilson}}, \bibinfo {author} {\bibfnamefont {A.~J.}\ \bibnamefont {Ribeiro}}, \ and\ \bibinfo {author} {\bibfnamefont {B.~L.}\ \bibnamefont {Pruitt}},\ }\href {\doibase 10.1371/JOURNAL.PONE.0189901} {\bibfield  {journal} {\bibinfo  {journal} {PLOS ONE}\ }\textbf {\bibinfo {volume} {13}},\ \bibinfo {pages} {e0189901} (\bibinfo {year} {2018})}\BibitemShut {NoStop}%
\bibitem [{\citenamefont {Mizutani}\ \emph {et~al.}(2009)\citenamefont {Mizutani}, \citenamefont {Kawabata}, \citenamefont {Koyama}, \citenamefont {Takahashi},\ and\ \citenamefont {Haga}}]{Mizutani2009RegulationCascade}%
  \BibitemOpen
  \bibfield  {author} {\bibinfo {author} {\bibfnamefont {T.}~\bibnamefont {Mizutani}}, \bibinfo {author} {\bibfnamefont {K.}~\bibnamefont {Kawabata}}, \bibinfo {author} {\bibfnamefont {Y.}~\bibnamefont {Koyama}}, \bibinfo {author} {\bibfnamefont {M.}~\bibnamefont {Takahashi}}, \ and\ \bibinfo {author} {\bibfnamefont {H.}~\bibnamefont {Haga}},\ }\href {\doibase 10.1002/CM.20378} {\bibfield  {journal} {\bibinfo  {journal} {Cell Motility and the Cytoskeleton}\ }\textbf {\bibinfo {volume} {66}},\ \bibinfo {pages} {389} (\bibinfo {year} {2009})}\BibitemShut {NoStop}%
\bibitem [{\citenamefont {Takemoto}\ \emph {et~al.}(2015)\citenamefont {Takemoto}, \citenamefont {Ishihara}, \citenamefont {Mizutani}, \citenamefont {Kawabata},\ and\ \citenamefont {Haga}}]{Takemoto2015CompressivePathway}%
  \BibitemOpen
  \bibfield  {author} {\bibinfo {author} {\bibfnamefont {K.}~\bibnamefont {Takemoto}}, \bibinfo {author} {\bibfnamefont {S.}~\bibnamefont {Ishihara}}, \bibinfo {author} {\bibfnamefont {T.}~\bibnamefont {Mizutani}}, \bibinfo {author} {\bibfnamefont {K.}~\bibnamefont {Kawabata}}, \ and\ \bibinfo {author} {\bibfnamefont {H.}~\bibnamefont {Haga}},\ }\href {\doibase 10.1371/JOURNAL.PONE.0117937} {\bibfield  {journal} {\bibinfo  {journal} {PloS one}\ }\textbf {\bibinfo {volume} {10}} (\bibinfo {year} {2015}),\ 10.1371/JOURNAL.PONE.0117937}\BibitemShut {NoStop}%
\bibitem [{\citenamefont {Noll}\ \emph {et~al.}(2017)\citenamefont {Noll}, \citenamefont {Mani}, \citenamefont {Heemskerk}, \citenamefont {Streichan},\ and\ \citenamefont {Shraiman}}]{Noll2017ActiveTissues}%
  \BibitemOpen
  \bibfield  {author} {\bibinfo {author} {\bibfnamefont {N.}~\bibnamefont {Noll}}, \bibinfo {author} {\bibfnamefont {M.}~\bibnamefont {Mani}}, \bibinfo {author} {\bibfnamefont {I.}~\bibnamefont {Heemskerk}}, \bibinfo {author} {\bibfnamefont {S.~J.}\ \bibnamefont {Streichan}}, \ and\ \bibinfo {author} {\bibfnamefont {B.~I.}\ \bibnamefont {Shraiman}},\ }\href {\doibase 10.1038/nphys4219} {\bibfield  {journal} {\bibinfo  {journal} {Nature Physics 2017 13:12}\ }\textbf {\bibinfo {volume} {13}},\ \bibinfo {pages} {1221} (\bibinfo {year} {2017})}\BibitemShut {NoStop}%
\bibitem [{\citenamefont {Mu{\~{n}}oz}\ and\ \citenamefont {Albo}(2013)}]{Munoz2013Physiology-basedViscoelasticity}%
  \BibitemOpen
  \bibfield  {author} {\bibinfo {author} {\bibfnamefont {J.~J.}\ \bibnamefont {Mu{\~{n}}oz}}\ and\ \bibinfo {author} {\bibfnamefont {S.}~\bibnamefont {Albo}},\ }\href {\doibase 10.1103/PhysRevE.88.012708} {\bibfield  {journal} {\bibinfo  {journal} {PHYSICAL REVIEW E}\ }\textbf {\bibinfo {volume} {88}},\ \bibinfo {pages} {12708} (\bibinfo {year} {2013})}\BibitemShut {NoStop}%
\bibitem [{\citenamefont {Hino}\ \emph {et~al.}(2020)\citenamefont {Hino}, \citenamefont {Rossetti}, \citenamefont {Mar{\'{i}}n-Llaurad{\'{o}}}, \citenamefont {Aoki}, \citenamefont {Trepat}, \citenamefont {Matsuda},\ and\ \citenamefont {Hirashima}}]{Hino2020ERK-MediatedPolarization}%
  \BibitemOpen
  \bibfield  {author} {\bibinfo {author} {\bibfnamefont {N.}~\bibnamefont {Hino}}, \bibinfo {author} {\bibfnamefont {L.}~\bibnamefont {Rossetti}}, \bibinfo {author} {\bibfnamefont {A.}~\bibnamefont {Mar{\'{i}}n-Llaurad{\'{o}}}}, \bibinfo {author} {\bibfnamefont {K.}~\bibnamefont {Aoki}}, \bibinfo {author} {\bibfnamefont {X.}~\bibnamefont {Trepat}}, \bibinfo {author} {\bibfnamefont {M.}~\bibnamefont {Matsuda}}, \ and\ \bibinfo {author} {\bibfnamefont {T.}~\bibnamefont {Hirashima}},\ }\href {\doibase 10.1016/J.DEVCEL.2020.05.011/ATTACHMENT/DD5E7D00-CE92-4AF3-BE09-1F3C7CD23975/MMC11.MP4} {\bibfield  {journal} {\bibinfo  {journal} {Developmental Cell}\ }\textbf {\bibinfo {volume} {53}},\ \bibinfo {pages} {646} (\bibinfo {year} {2020})}\BibitemShut {NoStop}%
\bibitem [{\citenamefont {Esfahani}\ \emph {et~al.}(2021)\citenamefont {Esfahani}, \citenamefont {Rosenbohm}, \citenamefont {Safa}, \citenamefont {Lavrik}, \citenamefont {Minnick}, \citenamefont {Zhou}, \citenamefont {Kong}, \citenamefont {Jin}, \citenamefont {Kim}, \citenamefont {Liu}, \citenamefont {Lu}, \citenamefont {Lim}, \citenamefont {Wahl}, \citenamefont {Dao}, \citenamefont {Huang},\ and\ \citenamefont {Yang}}]{Esfahani2021CharacterizationJunctions}%
  \BibitemOpen
  \bibfield  {author} {\bibinfo {author} {\bibfnamefont {A.~M.}\ \bibnamefont {Esfahani}}, \bibinfo {author} {\bibfnamefont {J.}~\bibnamefont {Rosenbohm}}, \bibinfo {author} {\bibfnamefont {B.~T.}\ \bibnamefont {Safa}}, \bibinfo {author} {\bibfnamefont {N.~V.}\ \bibnamefont {Lavrik}}, \bibinfo {author} {\bibfnamefont {G.}~\bibnamefont {Minnick}}, \bibinfo {author} {\bibfnamefont {Q.}~\bibnamefont {Zhou}}, \bibinfo {author} {\bibfnamefont {F.}~\bibnamefont {Kong}}, \bibinfo {author} {\bibfnamefont {X.}~\bibnamefont {Jin}}, \bibinfo {author} {\bibfnamefont {E.}~\bibnamefont {Kim}}, \bibinfo {author} {\bibfnamefont {Y.}~\bibnamefont {Liu}}, \bibinfo {author} {\bibfnamefont {Y.}~\bibnamefont {Lu}}, \bibinfo {author} {\bibfnamefont {J.~Y.}\ \bibnamefont {Lim}}, \bibinfo {author} {\bibfnamefont {J.~K.}\ \bibnamefont {Wahl}}, \bibinfo {author} {\bibfnamefont {M.}~\bibnamefont {Dao}}, \bibinfo {author} {\bibfnamefont {C.}~\bibnamefont {Huang}}, \ and\ \bibinfo {author} {\bibfnamefont {R.}~\bibnamefont {Yang}},\
  }\href {\doibase 10.1073/PNAS.2019347118/SUPPL{\_}FILE/PNAS.2019347118.SM04.MP4} {\bibfield  {journal} {\bibinfo  {journal} {Proceedings of the National Academy of Sciences of the United States of America}\ }\textbf {\bibinfo {volume} {118}},\ \bibinfo {pages} {e2019347118} (\bibinfo {year} {2021})}\BibitemShut {NoStop}%
\bibitem [{\citenamefont {Kollmannsberger}\ and\ \citenamefont {Fabry}(2011)}]{Kollmannsberger2011LinearCells}%
  \BibitemOpen
  \bibfield  {author} {\bibinfo {author} {\bibfnamefont {P.}~\bibnamefont {Kollmannsberger}}\ and\ \bibinfo {author} {\bibfnamefont {B.}~\bibnamefont {Fabry}},\ }\href {\doibase 10.1146/ANNUREV-MATSCI-062910-100351} {\bibfield  {journal} {\bibinfo  {journal} {Annual Review of Materials Research}\ }\textbf {\bibinfo {volume} {41}},\ \bibinfo {pages} {75} (\bibinfo {year} {2011})}\BibitemShut {NoStop}%
\bibitem [{\citenamefont {Mu{\~{n}}oz}\ \emph {et~al.}(2018)\citenamefont {Mu{\~{n}}oz}, \citenamefont {Dingle},\ and\ \citenamefont {Wenzel}}]{Munoz2018MechanicalChanges}%
  \BibitemOpen
  \bibfield  {author} {\bibinfo {author} {\bibfnamefont {J.~J.}\ \bibnamefont {Mu{\~{n}}oz}}, \bibinfo {author} {\bibfnamefont {M.}~\bibnamefont {Dingle}}, \ and\ \bibinfo {author} {\bibfnamefont {M.}~\bibnamefont {Wenzel}},\ }\href {\doibase 10.1103/PHYSREVE.98.052409/FIGURES/9/MEDIUM} {\bibfield  {journal} {\bibinfo  {journal} {Physical Review E}\ }\textbf {\bibinfo {volume} {98}},\ \bibinfo {pages} {052409} (\bibinfo {year} {2018})}\BibitemShut {NoStop}%
\bibitem [{\citenamefont {Ron}\ \emph {et~al.}(2020)\citenamefont {Ron}, \citenamefont {Monzo}, \citenamefont {Gauthier}, \citenamefont {Voituriez},\ and\ \citenamefont {Gov}}]{Ron2020One-dimensionalPatterns}%
  \BibitemOpen
  \bibfield  {author} {\bibinfo {author} {\bibfnamefont {J.~E.}\ \bibnamefont {Ron}}, \bibinfo {author} {\bibfnamefont {P.}~\bibnamefont {Monzo}}, \bibinfo {author} {\bibfnamefont {N.~C.}\ \bibnamefont {Gauthier}}, \bibinfo {author} {\bibfnamefont {R.}~\bibnamefont {Voituriez}}, \ and\ \bibinfo {author} {\bibfnamefont {N.~S.}\ \bibnamefont {Gov}},\ }\href {\doibase 10.1103/PHYSREVRESEARCH.2.033237/FIGURES/28/MEDIUM} {\bibfield  {journal} {\bibinfo  {journal} {Physical Review Research}\ }\textbf {\bibinfo {volume} {2}},\ \bibinfo {pages} {033237} (\bibinfo {year} {2020})}\BibitemShut {NoStop}%
\bibitem [{\citenamefont {Lo~Vecchio}\ \emph {et~al.}(2024)\citenamefont {Lo~Vecchio}, \citenamefont {Pertz}, \citenamefont {Szopos}, \citenamefont {Navoret},\ and\ \citenamefont {Riveline}}]{LoVecchio2024SpontaneousBoundaries}%
  \BibitemOpen
  \bibfield  {author} {\bibinfo {author} {\bibfnamefont {S.}~\bibnamefont {Lo~Vecchio}}, \bibinfo {author} {\bibfnamefont {O.}~\bibnamefont {Pertz}}, \bibinfo {author} {\bibfnamefont {M.}~\bibnamefont {Szopos}}, \bibinfo {author} {\bibfnamefont {L.}~\bibnamefont {Navoret}}, \ and\ \bibinfo {author} {\bibfnamefont {D.}~\bibnamefont {Riveline}},\ }\href {\doibase 10.1038/s41567-023-02295-x} {\bibfield  {journal} {\bibinfo  {journal} {Nature Physics 2024}\ ,\ \bibinfo {pages} {1}} (\bibinfo {year} {2024})}\BibitemShut {NoStop}%
\bibitem [{\citenamefont {Dow}(2023)}]{Dow2023UCContacts}%
  \BibitemOpen
  \bibfield  {author} {\bibinfo {author} {\bibfnamefont {L.}~\bibnamefont {Dow}},\ }\emph {\bibinfo {title} {https://escholarship.org/uc/item/6j31f1wk}},\ \href {https://escholarship.org/uc/item/6j31f1wk} {Ph.D. thesis},\ \bibinfo  {school} {University of California, Santa Barbara} (\bibinfo {year} {2023})\BibitemShut {NoStop}%
\bibitem [{\citenamefont {Jain}\ \emph {et~al.}(2020)\citenamefont {Jain}, \citenamefont {Cachoux}, \citenamefont {Narayana}, \citenamefont {de~Beco}, \citenamefont {D’Alessandro}, \citenamefont {Cellerin}, \citenamefont {Chen}, \citenamefont {Heuz{\'{e}}}, \citenamefont {Marcq}, \citenamefont {M{\`{e}}ge}, \citenamefont {Kabla}, \citenamefont {Lim},\ and\ \citenamefont {Ladoux}}]{Jain2020}%
  \BibitemOpen
  \bibfield  {author} {\bibinfo {author} {\bibfnamefont {S.}~\bibnamefont {Jain}}, \bibinfo {author} {\bibfnamefont {V.~M.}\ \bibnamefont {Cachoux}}, \bibinfo {author} {\bibfnamefont {G.~H.}\ \bibnamefont {Narayana}}, \bibinfo {author} {\bibfnamefont {S.}~\bibnamefont {de~Beco}}, \bibinfo {author} {\bibfnamefont {J.}~\bibnamefont {D’Alessandro}}, \bibinfo {author} {\bibfnamefont {V.}~\bibnamefont {Cellerin}}, \bibinfo {author} {\bibfnamefont {T.}~\bibnamefont {Chen}}, \bibinfo {author} {\bibfnamefont {M.~L.}\ \bibnamefont {Heuz{\'{e}}}}, \bibinfo {author} {\bibfnamefont {P.}~\bibnamefont {Marcq}}, \bibinfo {author} {\bibfnamefont {R.~M.}\ \bibnamefont {M{\`{e}}ge}}, \bibinfo {author} {\bibfnamefont {A.~J.}\ \bibnamefont {Kabla}}, \bibinfo {author} {\bibfnamefont {C.~T.}\ \bibnamefont {Lim}}, \ and\ \bibinfo {author} {\bibfnamefont {B.}~\bibnamefont {Ladoux}},\ }\href {\doibase 10.1038/s41567-020-0875-z} {\bibfield  {journal} {\bibinfo  {journal} {Nature Physics 2020 16:7}\ }\textbf {\bibinfo {volume}
  {16}},\ \bibinfo {pages} {802} (\bibinfo {year} {2020})}\BibitemShut {NoStop}%
\end{thebibliography}%

\end{document}